\documentclass[useAMS,usenatbib]{mn2e_mod}
\usepackage{natbib}
\usepackage{graphicx}
\usepackage{amsmath}
\usepackage{amssymb}
\usepackage{deluxetable}
\usepackage{longtable}
\usepackage{url}

\newcommand{\cloudy}{{\sc cloudy}}

\let\ACMmaketitle=\maketitle
\renewcommand{\maketitle}{\begingroup\let\footnote=\thanks \ACMmaketitle\endgroup}

\title[outflow rates and energetics]{Discovering AGN-driven winds through their infrared emission: II. Mass outflow rate and energetics}

\author[Baron \& Netzer]
{Dalya Baron$^{1}$\thanks{dalyabaron@gmail.com} \&
Hagai Netzer$^{1}$
\\
\\
$^{1}$School of Physics and Astronomy, Tel-Aviv University, Tel Aviv 69978, Israel.\\
}

\begin{document}

\maketitle

\label{firstpage}
\begin{abstract}

The global influence of AGN-driven outflows remains uncertain, due to a lack of large samples with accurately-determined outflow properties. In the second paper of this series, we determine the mass and energetics of ionized outflows is 234 type II AGN, the largest such sample to date, by combining the infrared emission of the dust in the wind (paper I) with the emission line properties. We provide new general expressions for the properties of the outflowing gas, which depend on the ionization state of the gas. We also present a novel method to estimate the electron density in the outflow, based on optical line ratios and on the known location of the wind. The inferred electron densities, $n_{\mathrm{e}} \sim 10^{4.5}\,\mathrm{cm^{-3}}$, are two orders of magnitude larger than typically found in most other cases of ionized outflows. We argue that the discrepancy is due to the fact that the commonly-used [SII]-based method underestimates the true density by a large factor. As a result, the inferred mass outflow rates and kinetic coupling efficiencies are $\dot{M}_{\mathrm{out}} \sim 10^{-2}\, \mathrm{(M_{\odot}/yr)}$ and $\epsilon = \dot{E}_{\mathrm{kin}}/L_{\mathrm{bol}} \sim 10^{-5}$ respectively, 1--2 orders of magnitude lower than previous estimates. Our analysis suggests the existence of a significant amount of neutral atomic gas at the back of the outflowing ionized gas clouds, with mass that is a factor of a few larger than the observed ionized gas mass. This has significant implications for the estimated mass and energetics of such flows.

\end{abstract}


\vspace{1cm}
\section{Introduction}\label{s:intro}

The discovery of correlations between the masses of super massive black holes (SMBHs) and several properties of their host galaxy (stellar velocity dispersion, bulge mass, and bulge luminosity; e.g. \citealt{gebhardt00a, ferrarese00, tremaine02, gultekin09}) have led to a suggestion that the growth of the SMBH is linked to the stellar mass growth in its host galaxy (e.g., \citealt{silk98, kauffmann00, zubovas14}). Active galactic nuclei (AGN) feedback, in the form of powerful outflows, is invoked as a way to couple the energy released by the accreting SMBH with the ISM of its host galaxy, providing a possible explanation for the observed correlations (e.g., \citealt{silk98,fabian99,king03,dimatteo05,springel05b}). A number of such models have successfully reproduced these correlations, by requiring that a significant amount of the accretion energy of the AGN will be mechanically-coupled to the ISM of the host galaxy ($\sim $5--10\% $L_{\mathrm{bol}}$; e.g., \citealt{fabian99,tremaine02, dimatteo05,springel05b,kurosawa09}). 

Powerfull winds are routinely detected in the host galaxies of AGN, spanning a large range of evolutionary stages, luminosities, and outflow gas phases (e.g., \citealt{nesvadba06, mullaney13, rupke13, veilleux13, cicone14, harrison14, cheung16, zakamska16, baron17b, fiore17, rupke17, baron18}). In order to assess the effect of such flows on their host galaxy evolution, it is necessary to accurately determine their mass and energetics, and compare them to those required by feedback models (see however \citealt{harrison18}). A major source of uncertainty is related to the relative contributions of AGN and supernovae to the observed outflows (see introduction in \citealt{baron18}). It is normally assumed that the source that photoionizes the stationary and outflowing gas is also the main driver of the observed winds (e.g., \citealt{karouzos16a, karouzos16b, forster_schreiber18}). Thus, outflows that are observed in systems with emission line ratios that are consistent with AGN photoionization, are considered to be driven by the AGN. 

A large number of AGN-driven winds are observed in the warm ionized phase, traced by rest-frame optical emission lines (e.g., \citealt{mullaney13, harrison14, karouzos16a, karouzos16b, forster_schreiber18}). In order to properly quantify the mass and energetics of these outflows, it is necessary to determine their kinematics (e.g., outflow velocity, $v_{out}$), spatial extent ($r_{out}$), and electron density ($n_{e}$), all of which are subjected to various uncertainties and systematics (see e.g., \citealt{husemann16, villar_martin16, harrison18, rose18, tadhunter18}). For example, the electron densities in such outflows are claimed to be rather low, in the range $\sim 10^{2} - 10^{3}\,\mathrm{cm^{-3}}$, based on the [SII]~$\lambda \lambda$ 6717,6731\AA\, doublet ratio (e.g., \citealt{harrison14, fiore17, karouzos16b}). However, recent studies suggest that the densities are higher, around $\sim 10^{3} - 10^{5}\,\mathrm{cm^{-3}}$, a regime where the [SII]-based method cannot be used \citep{holt11, rose18, santoro18, spence18}. 

So far, only spatially-resolved spectroscopic observations, either by long-spit observations or using integral field units (IFUs), could be used to determine the outflow extent, $r_{out}$, and thus determine the mass and energetics of outflows (e.g., \citealt{sharp10, fischer11, liu13a, liu13b, harrison14, karouzos16a, karouzos16b, baron18, perna19}). While such observations offer a more detailed view of the outflows, they are observationally demanding. Typically, only the most extreme outflow cases are followed-up with IFUs (e.g., \citealt{mullaney13, harrison14}), resulting in biased samples that might not represent more typical outflow cases. Furthermore, IFU-based observations suffer from various uncertainties and systematics, such as projection effects and beam smearing, that can significantly affect the derived kinematics and extents of the outflows (e.g., \citealt{husemann16, villar_martin16, fischer18, tadhunter18}).

In \citet[hereafter Paper I]{baron19}, we argued that the outflowing gas in active galaxies contains dust, which is heated by the central AGN, and emits in mid-infrared wavelengths. We examined the infrared spectral energy distribution (SED) of thousands of type II AGN, and showed that emission by such dust is detected in many systems that host ionized gas outflows. Our SED fitting provides the properties of the dust, including its distance from the central source. Since this dust is mixed with the outflowing gas, our method provides the luminosity-weighted (and therefore the mass-weighted) location of the outflow. This infrared emission is not subjected to various systematics affecting ground-based optical IFU observations, such as beam smearing, projection effects, and dust extinction that affects the receding part of the outflow. The estimated location of the dust, $r_{\mathrm{dust}}$, can be combined with 1D spectroscopic observations to determine the mass and energetics of ionized outflows in hundreds of systems. 

In this work we focus on a subset of the sample presented in Paper I, in which ionized outflows are detected in many optical emission lines ([OIII], H$\beta$, [OI], [NII], H$\alpha$, and [SII]), and there is a clear detection of a dusty wind component at mid-infrared wavelengths. This combination allows us to accurately determine key properties of the gas in the wind, such as dust reddening, ionization state, and electron density, and thus constrain the mass and energetics of the winds. We describe our sample in section \ref{s:sample}, and derive the spectral properties of the objects in section \ref{s:spectral_props}. We present new general expressions to derive the ionization parameter, electron density, and line emissivity of the ionized gas in section \ref{s:photo_estimates}. We then estimate the outflowing gas mass, mass outflow rate, and kinetic energy of the winds in section \ref{s:mass_and_energy}. We discuss our results in section \ref{s:disc}, and conclude in section \ref{s:concs}.

\section{Sample Selection}\label{s:sample}

The sample analyzed in this paper has been described in detail in paper I. It consists of type II AGN with spectroscopically-detected outflows, for which a dusty wind component is detected in mid-infrared wavelengths.  For clarity, we summarize here the main properties of the sample.

We start with the publicly-available catalog: AGN Line Profile And Kinematics Archive (ALPAKA; \citealt{mullaney13}), which provides emission line measurements for a sample of 24\,264 optically-selected AGN from SDSS DR7 \citep{abazajian09}. \citet{mullaney13} performed a multi-component fitting to the emission lines [OIII]$\lambda$5007\AA, [NII]$\lambda$6584\AA, H$\alpha$, and H$\beta$, using continuum-subtracted spectra. The fit included one to three kinematic components to each of the emission lines, and the presence of an additional broader component in [OIII]$\lambda$5007\AA\, was interpreted as a spectroscopic signture of an ionized outflow. We selected systems that are classified in the ALPAKA catalog as type II AGN in the redshift range $0.05 \leq z \leq 0.15$, with spectroscopically-detected outflows.

We cross-matched this sample with the MPA-JHU catalog, which provides stellar mass and star formation rate (SFR) measurements of SDSS galaxies \citep{b04, kauff03b, t04, salim07}. We also used optical and infrared (IR) photometric data as follows. We used the $riz$ optical photometry by the SDSS, the $JHK$ near-infrared photometry from the 2MASS All-Sky Extended Source Catalog \citep{skrutskie06}, and the W1-W4 mid-infrared photometry from WISE \citep{wright10}. We used the astrometric cross-matches provided by the SDSS\footnote{https://skyserver.sdss.org/CasJobs} to cross-match between systems observed by SDSS and systems observed by 2MASS and WISE.

In Paper I, we examined the infrared SEDs of 2\,377 systems that show spectroscopic signatures of an ionized wind. The best-fitting SED included contributions from direct stellar light, which dominates the SED at optical and near-infrared wavelengths, torus and NLR dust emission, which dominate the SED at mid-infrared wavelengths, and dust in star forming (SF) regions which dominates the far-infrared wavelengths. The IR SED also included a new contribution from dust that is mixed with the outflow, is heated by the AGN, and emits at mid-infrared wavelengths. We found that 2\,044 systems require the additional dusty wind component. Out of these, we were able to constrain the dust temperature, covering factor, and location in 1\,696 systems. The initial sub-sample addressed in this paper consists of these 1\,696 type II AGN. 

\begin{figure}
\includegraphics[width=3.25in]{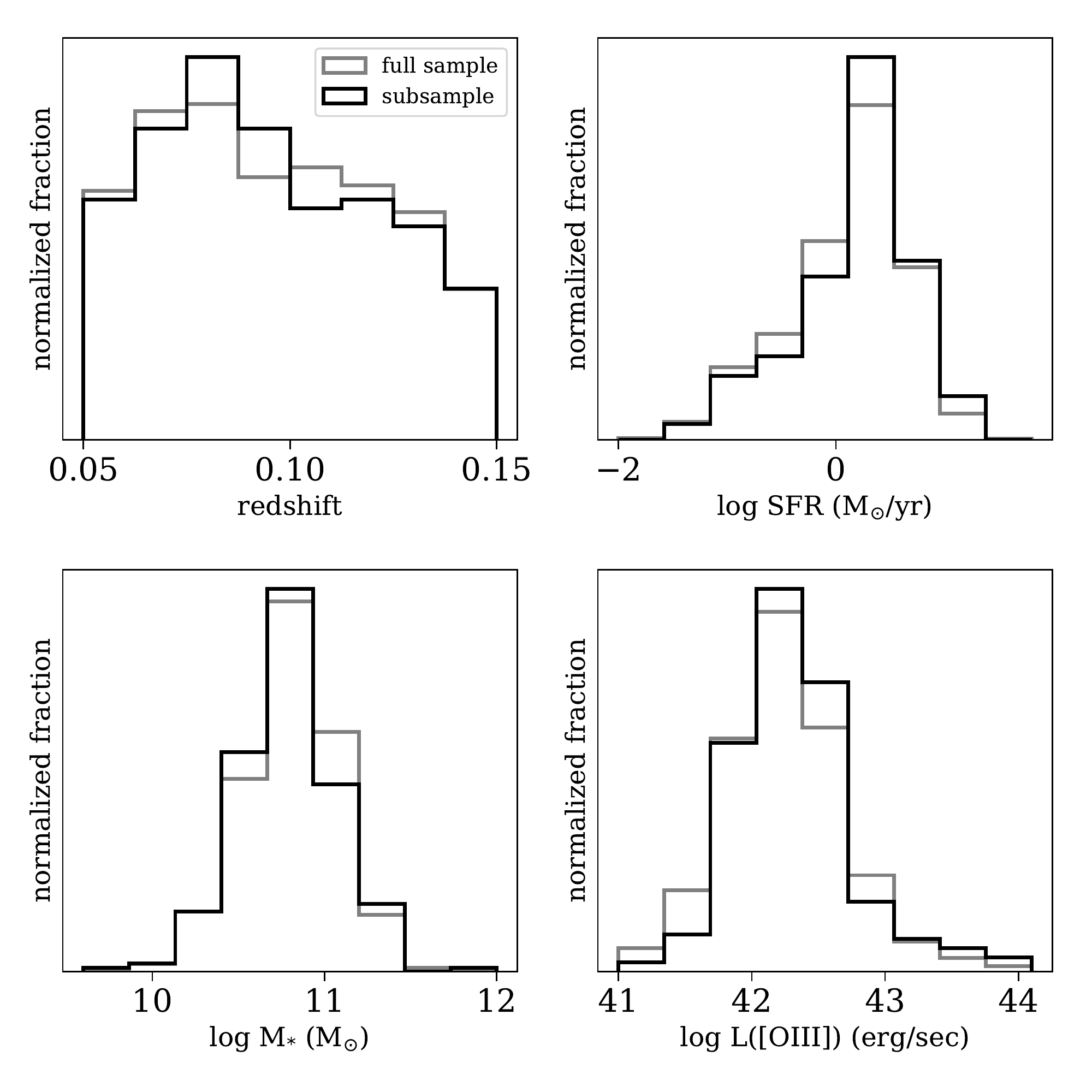}
\caption{Properties of the 234 type II AGN sample with a $2 \sigma$ detection of broad component in [OIII], [NII], [SII], H$\alpha$, and H$\beta$, and a detection of a dusty wind component at mid-infrared wavelengths, with well determined temperature, covering factor, and location. The SFR and stellar mass are taken from the MPA-JHU catalog, and the redshift and dust-corrected narrow [OIII] luminosity are taken from the ALPAKA catalog. The distributions of the subsample (black histogram) are roughly similar to the distribution of properties in our parent sample of 1696 objects (grey histogram).}\label{f:initial_sample_histograms}
\end{figure}

\section{Spectral properties}\label{s:spectral_props}

The goal in paper II is to determine the level of ionization, the density, the mass outflow rate and the kinetic energy of the outflowing gas. To study the ionization state and density of the gas, we must detect the outflow in various emission lines: [OIII], [NII], [SII], [OI], H$\alpha$, and H$\beta$ (see e.g., \citealt{baron17b}). The outflowing gas is traced by \emph{broader} emission lines that are blended with narrower emission lines coming from the stationary NLR. The broader emission lines are usually weaker than their corresponding narrow lines, and their characterization requires a careful emission line decomposition procedure. The ALPAKA catalog provides such a decomposition. However, \citet{mullaney13} fit the continuum emission locally using a 5 degree polynomial, which they then subtract to obtain an emission line spectrum. Since they do not fit the spectra with a stellar population synthesis model, stellar absorption lines are not properly accounted for, and may affect the resulting emission line spectrum, particularly around the H$\alpha$ and H$\beta$ emission lines. The broader H$\beta$ line is usually the weakest emission line in the spectrum, and its equivalent width is of the same order as the stellar H$\beta$ absorption. 

To improve the accuracy of the broad H$\beta$ emission line, we fit stellar population synthesis models to the 1\,696 type II AGN sample, and performed our own emission line decomposition. We use Penalized Pixel-Fitting stellar kinematics extraction code (pPXF; \citealt{cappellari12}), which is a public code for extracting the stellar kinematics and stellar population from absorption-line spectra of galaxies \citep{cappellari04}. It uses the MILES library, which contains single stellar population (SSP) synthesis models and covers the full range of the optical spectrum with a resolution of full width at half maximum (FWHM) of 2.3\AA\, \citep{vazdekis10}. We use SSP models with stellar ages that range from 0.03 to 14 Gyr, thus ensuring that we can properly describe spectra of systems with different star formation histories. The wide wavelength range of the SDSS spectra and the SSP models includes the Balmer absorption lines: H$\zeta$, H$\epsilon$, H$\delta$, H$\gamma$, H$\beta$, and H$\alpha$, ensuring that the fitted stellar continuum around the H$\beta$ absorption line is robust. The output of the code includes the relative weight of stars with different ages, the stellar velocity dispersion, the dust reddening towards the stars, and the best-fitting stellar model. Since we are not interested in the stellar properties of the galaxies in our sample, we only use the best-fitting stellar models and subtract them from the spectra.

The subtracted emission line spectra show varios emission lines: [OIII]~$\lambda \lambda$ 4959,5007\AA, $\mathrm{H\alpha}$~$\lambda$ 6563\AA, [NII]~$\lambda \lambda$ 6548,6584\AA, [OII]~$\lambda \lambda$ 3725,3727\AA, $\mathrm{H\beta}$~$\lambda$ 4861\AA, [OI]~$\lambda \lambda$ 6300,6363\AA, and [SII]~$\lambda \lambda$ 6717,6731\AA\, (hereafter $\mathrm{[OIII]}$, $\mathrm{H\alpha}$, $\mathrm{[NII]}$, $\mathrm{[OII]}$, $\mathrm{H\beta}$, $\mathrm{[OI]}$, and $\mathrm{[SII]}$). We model each emission line as a sum of two Gaussians - one which represents the narrow component and one that represents the broader, outflowing, component. We perform a joint fit to all the emission lines under several constrains: (1) we force the intensity ratio of the emission lines [OIII]~$\lambda \lambda$ 4959,5007\AA\, and [NII]~$\lambda \lambda$ 6548,6584\AA\, to the theoretical ratio of 1:3, (2) we tie the central wavelengths of all the narrow and broad lines so that the gas has the same systemic velocity, and (3) we force the widths of all the narrow lines to show the same velocity dispersion, and we do the same for the broader lines. 

\begin{figure}
\includegraphics[width=3.25in]{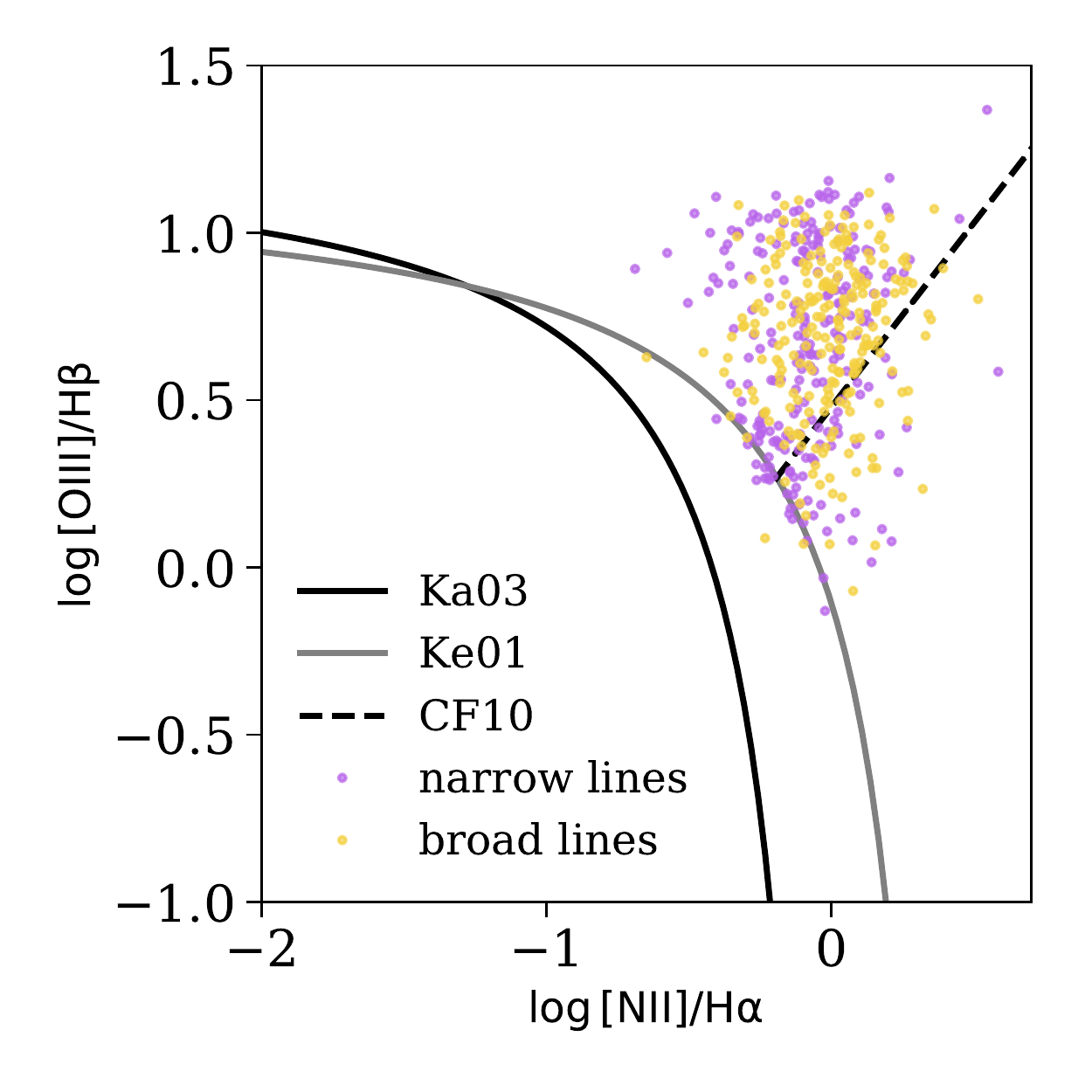}
\caption{[OIII]/H$\beta$ versus [NII]/H$\alpha$ for the narrow (purple) and broad (yellow) emission lines in our sample of 234 type-II AGN. We mark the extreme starburst line of Ke01 with black, the composite line of Ka03 with grey, and the LINER-Seyfert separation line of CF10 with a dashed black line.}\label{f:bpt_diagram}
\end{figure}

\begin{figure*}
\includegraphics[width=0.9\textwidth]{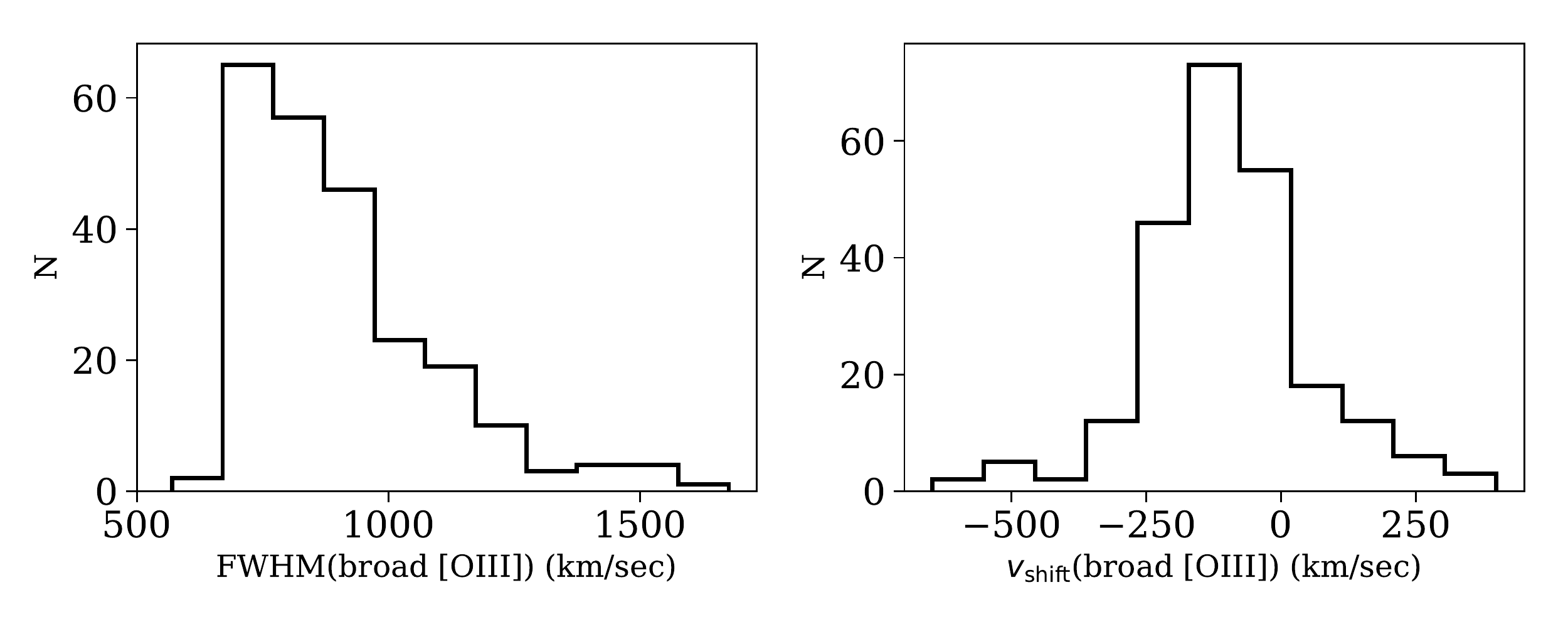}
\caption{\textbf{Left panel:} the distribution of the FWHM of the broad emission lines in our sample. \textbf{Right panel:} the distribution of the velocity shift of the broad emission lines with respect to the narrow emission lines. The shift is defined as the difference between the centroids of the lines. By construction, it is the same for the different emission lines.}\label{f:broad_lines_kinematics}
\end{figure*}

Out of our initial sample, we detect a broad component with at least 2$\sigma$ in [OIII], [NII], [SII], H$\alpha$, and H$\beta$, in 234 systems. We focus on these 234 systems throughout the paper. In figure \ref{f:initial_sample_histograms} we show the distribution of redshift, SFR, stellar mass, and dust-corrected narrow [OIII] luminosity for this subsample. The distributions are roughly similar to the distributions of these properties in the parent sample of 1\,696 type II AGN. The two-sample KS test for the distributions give p-values of 0.53, 0.04, 0.50, and 0.03, for the redshift, SFR, stellar mass, and [OIII] luminosity respectively.

In figure \ref{f:bpt_diagram} we show the narrow and broad emission lines on a line-diagnostic diagram \citep{baldwin81, veilleux87}. We show three separating criteria. The first is a theoretical upper limit which separates starbursts and AGN-dominated galaxies (\citealt{kewley01}, Ke01; black line). The second is a modified criterion which includes composite galaxies showing contributions from both SF and AGN (\citealt{kauff03a}, Ka03; grey line). The third is a line from \citet{cidfernandes10} to separate between LINERs and Seyferts (CF10; dashed black). Clearly, both the stationary (narrow lines) and the outflowing (broad lines) gas in most of the 234 systems are dominated by AGN photoionization, with the majority of the systems classified as Seyferts and some as LINERs. 

In the left panel of figure \ref{f:broad_lines_kinematics} we show the FWHM of the broad emission lines. Since the fit requires all the broad lines to have the same velocity dispersion, the FWHM is the same for all the broad lines. In the right panel of figure \ref{f:broad_lines_kinematics} we show the velocity shift of the broad emission lines with respect to the systemic velocity, defined by the centroid of the narrow emission lines. As in the left panel, this velocity shift is the same in all the lines.

\begin{figure*}
\includegraphics[width=0.9\textwidth]{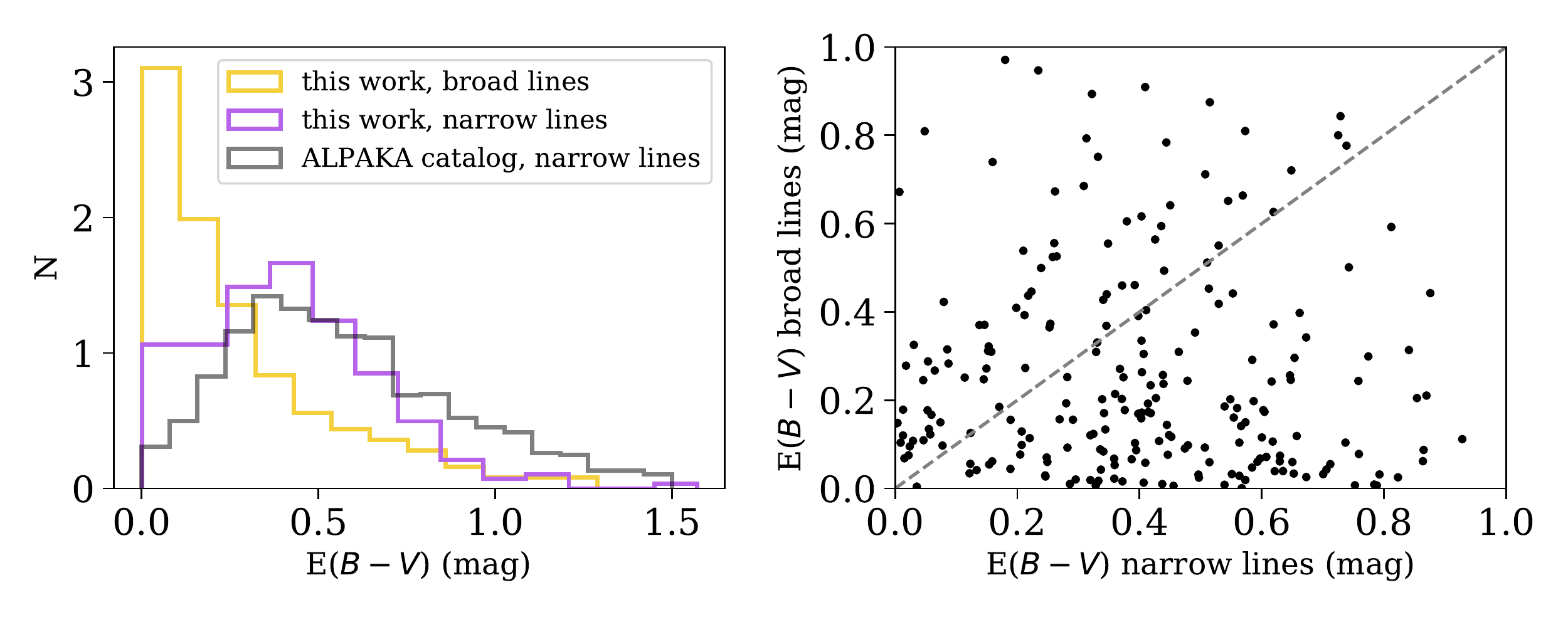}
\caption{\textbf{Left panel:} the distribution of colour excess towards the narrow lines (purple) and the broad lines (yellow) in our sample. For comparison, we show the distribution of the colour excess towards the narrow lines listed in the ALPAKA catalog (black). \textbf{Right panel:} colour excess towards the broad lines versus colour excess towards the narrow lines. }\label{f:dust_reddening_distribution}
\end{figure*}

Next, we use the measured H$\alpha$/H$\beta$ flux ratios to estimate the dust reddening towards the two kinematic components. Assuming case-B recombination, a gas temperature of $10^4$ K, a dusty screen, and the \citet[CCM]{cardelli89} extinction law, the colour excess is given by:
\begin{equation}\label{eq:1}
	{\mathrm{E}(B-V) = \mathrm{2.33 \times log\, \Bigg[ \frac{(H\alpha/H\beta)_{obs}}{2.85} \Bigg] }},
\end{equation}
where $\mathrm{(H\alpha/H\beta)_{obs}}$ is the observed line ratio. In the left panel of figure \ref{f:dust_reddening_distribution} we show the distribution of the colour excess for the narrow and broad lines respectively. For comparison, we show the distribution of the colour excess for the narrow emission lines in the ALPAKA catalog. Our narrow-line-based distribution of $\mathrm{E}(B-V)$, and the one from the ALPAKA catalog are roughly consistent, with somewhat larger number of sources with higher $\mathrm{E}(B-V)$ in the latter distribution. We attribute this difference to the different procedures used to fit and subtract the stellar continuum emission. For this reason, we do not compare the colour excess distributions in the broad emission lines.

In the right panel of figure \ref{f:dust_reddening_distribution} we show the colour excess towards the broad lines versus the colour excess towards the narrow lines. Surprisingly, we find no correlation between the two. This is consistent with the findings of \citet{rose18}, who analysed a sample of AGN-driven outflows in ultra luminous infrared galaxies. The figure clearly shows that the narrow and the broad lines suffer from different amounts of extinction (see also \citealt{baron17b, baron18}). Various earlier studies assume the same amount of dust extinction, and use the estimated colour excess in the narrow lines to correct the broad line luminosities for dust extinction. Figure \ref{f:dust_reddening_distribution} suggests that such a procedure is not justified. 

\begin{figure}
	\center
\includegraphics[width=3.in]{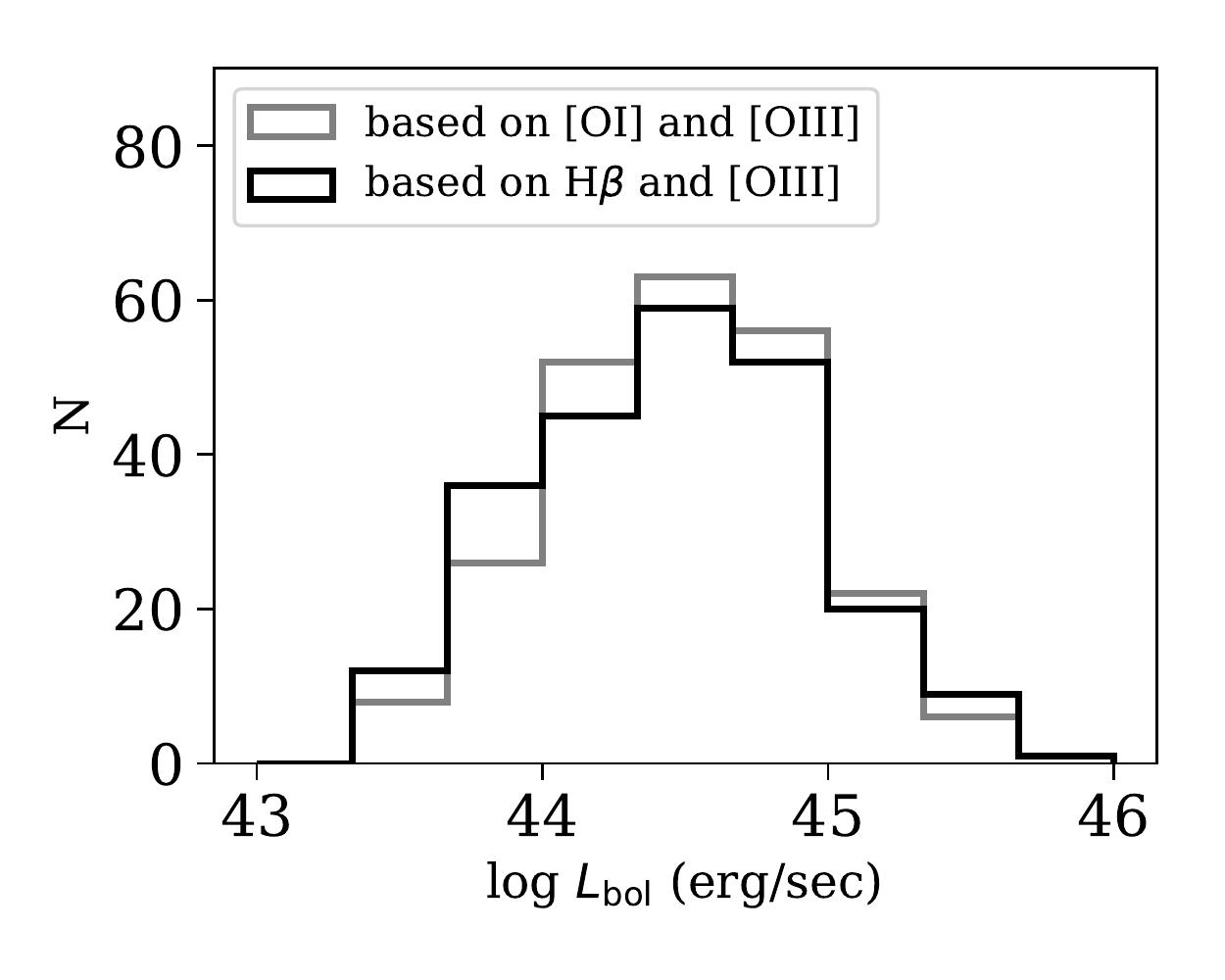}
\caption{The distribution of AGN bolometric luminosity in our sample using the two methods explained in the text.}\label{f:lbol_distribution}
\end{figure}

We derived dust-corrected line luminosities using the measured $\mathrm{E}(B-V)$. We propagate the uncertainties on the best-fitting parameters to obtain the uncertainties of the line luminosities, which are typically in the range 0.05--0.1 dex. We use the narrow emission line luminosities, which are based only on the narrow components from the multi-Gaussian fits, to estimate the bolometric luminosity of the AGN with two different methods. The first uses the narrow H$\beta$ and [OIII] luminosities (equation 1 in \citealt{netzer09}), and the second uses the narrow [OI] and [OIII] luminosities (equation 3 in \citealt{netzer09}). In figure \ref{f:lbol_distribution} we show the distribution of AGN bolometric luminosity in our sample. The two methods give consistent estimates. One can see that our sample spans about two orders of magnitude in AGN bolometric luminosity, roughly from $\log L_{\mathrm{bol}} = 43.5$ to $\log L_{\mathrm{bol}} = 45.5\, \mathrm{erg/sec}$.

\section{Gas properties from photoionization modelling}\label{s:photo_estimates}

In the next section we estimate the mass outflow rates and energetics of the observed outflows, which depend on various properties of the emission line gas, such as its electron density and level of ionization. Studies usually assume constant fiducial values for the electron densities and line emissivities in all the objects in the sample (see e.g., \citealt{harrison14, karouzos16a, fiore17}), which correspond to specific gas properties that might not be appropriate in a general case. In this section we provide more general expressions for the electron densities and line emissivities, which depend on the observed properties of individual sources. 

Many of the gas properties depend on the ionization parameter, $U = Q(\mathrm{Lyman})/ 4 \pi r^2 n_{\mathrm{H}} c$, where $Q(\mathrm{Lyman})$ is the number of the hydrogen-ionizing photons, $n_{\mathrm{H}}$ is the hydrogen density, and $c$ is the speed of light. In section \ref{s:ion_par_est} we show that the ionization parameter can be constrained using observed emission lines ratios. We then present a novel method to estimate the electron density in the gas, which is based on the estimated ionization parameter and on the known location of the gas, in section \ref{s:elec_dens_est}. Finally, we provide more general expressions of the H$\alpha$ and [OIII] line emissivities, where the latter depends on the estimated ionization parameter, in section \ref{s:line_emis}.

\subsection{Ionization parameter estimation}\label{s:ion_par_est}

The expressions we provide are based on extensive photoionization modelling, which details are listed in appendix \ref{a:photoionization_models}. We use a "single cloud" approach, where we calculate the properties of spherical shells of gas, with different density, pressure, column density, and distance from the central ionizing source. These models do not intend to replace more complex realistic models, that account for the geometrical distribution of clouds, with a range of densities and locations (see \citealt{netzer13} for a more complete description of multi-component photoionization models). Nevertheless, they provide useful expressions that can be used in the more realistic scenarios. 

We focus on models with optically-thick geometrically-thin shells of gas, and do not model cases in which the gas is optically-thin and extends over regions that are as large, or even larger, than the distance from the central source (e.g., see \citealt{baron18}). The latter models require information on the gas spatial distribution, which is not available in this case. In addition, the outflowing gas can be a combination of matter-bounded (optically-thin) and radiation-bounded (optically-thick) clouds (see e.g., \citealt{binette96}). We did not consider such a combination since (1) the optically-thick models we examine provide a general and consistent description of the four emission lines studied here ([OIII], [NII], H$\alpha$, and H$\beta$), and (2) the contribution of matter-bounded clouds to Balmer line luminosities, and to the outflowing dust emission, must be small unless the covering factor of this component is considerably larger than the one found here for radiation-bounded gas. This is inconsistent with our measurements, and thus the presence of such a component must have a small effect on the expressions derived here.

We examined a large range in gas metallicity (from 0.5$\mathrm{Z_{\odot}}$ to 2$\mathrm{Z_{\odot}}$), ionization parameter (from $\log U = -3.8$ to $\log U = -2$), and several different slopes of the ionizing continuum (with mean energy of an ionizing photon of 2.56 Ryd to 4.17 Ryd). We compared the predictions of constant-density versus constant-pressure models, and found consistent results within the range of properties we considered. Our analysis suggests that the expressions we provide are general, and do not depend on specific assumptions made during the modelling.

The [OIII]/H$\beta$ and [NII]/H$\alpha$ emission line ratios are based on the strongest lines in the optical spectra of AGN. These ratios are insensitive to dust reddening and are widely used to distinguish between AGN and SF as the main source of photoionization \citep{baldwin81, veilleux87, kewley01, kauff03a, cidfernandes10}. The [OIII]/H$\beta$ ratio is mostly sensitive to the ionization parameter in the gas, but it also depends on the gas metallicity. The [NII]/H$\alpha$ ratio is mostly sensitive to the gas metallicity, but it also depends on the ionization parameter. For AGN-dominated systems, both ratios depend slightly on the shape of the ionizing continuum (e.g., see figure \ref{f:all_models_BPT_diagram} and \citealt{groves04a, groves04b, kewley13}). 

We looked for an expression that ties the ionization parameter in the gas with the observed [OIII]/H$\beta$ and [NII]/H$\alpha$ line ratios. It is given by:
\begin{equation}\label{eq:5}
	\begin{split}
& \log U = a_{1} + a_{2}\Big[\log\mathrm{\Big(\frac{[OIII]}{H\beta}\Big)}\Big] + a_{3}\big[\log\mathrm{\Big(\frac{[OIII]}{H\beta}\Big)}\Big]^{2} + \\
& + a_{4}\Big[\log\mathrm{\Big(\frac{[NII]}{H\alpha}\Big)}\Big] + a_{5}\Big[\log\mathrm{\Big(\frac{[NII]}{H\alpha}\Big)}\Big]^{2},
	\end{split}
\end{equation}
where $a_{1}$, $a_{2}$, $a_{3}$, $a_{4}$, and $a_{5}$ are constants which are determined by a likelihood method. In appendix \ref{a:photoionization_models} we provide details of this fitting process and derive best fitting values for the five constants. These are listed in table \ref{t:ionization_parameter_coefficients}. Equation \ref{eq:5} is only valid for ionization parameters in the range $\log U = -2$ to $\log U = -3.8$. The typical deviation in $U$, given the range of parameters listed in the appendix, and the two types of models (constant density and constant total pressure) is 0.11 dex. This includes a large range of metallicities and several SED shapes. 

\begin{table}
	\centering 
	\tablewidth{0.8\linewidth} 
\begin{tabular}{|l|l|l|l|l|}
	\hline
 $a_{1}$ & $a_{2}$ & $a_{3}$ & $a_{4}$ & $a_{5}$\\
 \hline
 -3.766 & 0.191 & 0.778 & -0.251 & 0.342 \\
 \hline
\end{tabular}
\caption{Best-fitting numerical coefficients of equation \ref{eq:5}. Additional details are provided in the appendix.}
\label{t:ionization_parameter_coefficients}
\end{table}

\subsection{Electron density estimation}\label{s:elec_dens_est}

We use an independent method to estimate the electron density in the gas using optical emission lines. The method relies on the ionization parameter of the ionized gas, and the known location of the gas with respect to the central source. Using the definition of the ionization parameter, $U$, the electron density in the gas is given by:
\begin{equation}\label{eq:3}
	{n_{\mathrm{e}} \approx n_{\mathrm{H}} = \frac{Q(\mathrm{Lyman})}{4 \pi r^2 c U}},
\end{equation}
where $Q(\mathrm{Lyman})$ can be estimated from the AGN bolometric luminosity and the assumed SED (see equation \ref{eq:7} and related text in appendix \ref{a:photoionization_models}). For the typical AGN in our sample, and the assumed SEDs, equation \ref{eq:3} can be written as: 
\begin{equation}\label{eq:4}
	{n_{\mathrm{e}} \approx 3.2 \Big(\frac{L_{\mathrm{bol}}}{10^{45}\, \mathrm{erg/sec}}\Big) \Big( \frac{r}{1\,\mathrm{kpc}} \Big)^{-2} \Big(\frac{1}{U}\Big) \, \mathrm{cm^{-3}}       },
\end{equation}
where $L_{\mathrm{bol}}$ is the AGN bolometric luminosity and $r$ is the known distance of the gas from the central source. The ionization parameter is estimated using the observed emission line ratios (equation \ref{eq:5}). Equation \ref{eq:4} can be used to estimate the electron density of both stationary and outflowing gas, as long as the AGN is the main source of ionizing radiation.

\subsection{Line emissivities estimation}\label{s:line_emis}

In this section we provide more general expressions for the line emissivities in ionized gas. We focus on the H$\alpha$ and [OIII] emissivities, $\gamma_{\mathrm{H\alpha}}$ and $\gamma_{\mathrm{[OIII]}}$, as these emission lines are typically used to estimate mass outflow rates of ionized outflows. Both emissivities depend on the electron temperature in the gas, with $\gamma_{\mathrm{H\alpha}} \propto T_{e}^{-1}$ and $\gamma_{\mathrm{[OIII]}} \propto e^{-1/T_{e}}/\sqrt{T_{e}}$ (e.g., \citealt{draine11}). The electron temperature depends on the gas density, metallicity, and the ionization parameter. Our photoionization models suggest that the electron temperature in the ionized gas is in the range 10\,000 K to 14\,000 K, consistent with observations (e.g., \citealt{perna19}). We therefore select $T_{e}$ = 12\,000 K, resulting in H$\alpha$ emissivity of $\gamma = 3 \times 10^{-25}\, \mathrm{erg\, cm^{3}\, sec^{-1}}$. 

The [OIII] line emissivity is given by:
\begin{equation}\label{eq:5_2}
	\begin{split}
& \gamma_{\mathrm{[OIII]}} = C_{\mathrm{[OIII]}} \times h \nu_{\mathrm{[OIII]}} \times \frac{n(\mathrm{O^{+2}})}{n(\mathrm{O})} \times \frac{n(\mathrm{O})}{n(\mathrm{H})}, 
	\end{split}
\end{equation}
where $n_{\mathrm{e}}C_{\mathrm{[OIII]}}$ is the collisional excitation rate of [OIII], which depends on the electron temperature exponentially ($C_{\mathrm{[OIII]}} \propto e^{-1/T_{e}}/\sqrt{T_{e}}$; \citealt{draine11}), $h \nu_{\mathrm{[OIII]}}$ is the photon energy, $n(\mathrm{O^{+2}})/n(\mathrm{O})$ is the fraction of $\mathrm{O^{+2}}$ ions, and $n(\mathrm{O})/n(\mathrm{H})$ is the oxygen abundance relative to hydrogen. The emissivity depends on the temperature, primarily through $C_{\mathrm{[OIII]}}$, on the gas metallicity, which sets $n(\mathrm{O})/n(\mathrm{H})$, and on the ionization parameter, which affects $n(\mathrm{O^{+2}})/n(\mathrm{O})$. We note that different ionizing SEDs result in somewhat different $n(\mathrm{O^{+2}})/n(\mathrm{O})$ ratios, but this effect is of secondary importance.

\citet{cano12} provided an estimate for the [OIII] line emissivity which is based on equation \ref{eq:5_2}. They assumed that $n(\mathrm{O^{+2}})/n(\mathrm{O}) \sim 1$, which is a reasonable assumption for their specific system, but does not apply in the general case. In particular, this assumption is not valid for most of the objects in our sample. Furthermore, they estimated the emissivity assuming solar abundance of oxygen and did not take into account that a non-negligible fraction of the oxygen atoms are depleted onto dust grains. Assuming electron temperature of 12\,000 K, solar abundance of oxygen, $n(\mathrm{O})/n(\mathrm{H}) = 4.25\times 10^{-4}$, and a mean energy of an ionizing photon of 2.56 Ryd (model 2 SED; see appendix), the [OIII] line emissivity is given by:
\begin{equation}\label{eq:5_3}
\begin{split}
& \gamma_{\mathrm{[OIII]}} \, (\mathrm{erg\,cm^{3}\,sec^{-1}}) \approx \\
& \begin{cases}
	2.06 \times 10^{-24} \times \Big(\frac{n(\mathrm{O})/n(\mathrm{H})}{4.25\times 10^{-4}}\Big) \times f(\log U), & \text{dusty gas} \\
	3.45 \times 10^{-24} \times \Big(\frac{n(\mathrm{O})/n(\mathrm{H})}{4.25\times 10^{-4}}\Big) \times f(\log U), & \text{dustless gas} \\
\end{cases},
\end{split}
\end{equation}
where $f(\log U)$ represents the dependence on the ionization parameter, and is listed in table \ref{t:oiii_emis_dependence}. The ionization parameter is estimated from observed line ratios in equation \ref{eq:5}. The [OIII] emissivity for the dusty gas case is a factor of $\sim 0.6$ smaller than the emissivity for the dustless gas case. This is a direct result of the depletion of oxygen onto dust grains, where only 60\% of the oxygen atoms remain in the gas phase. In the next section, we use the [OIII] emissivity for the dusty gas case. Finally, we note that the constants in equation \ref{eq:5_3} represent the average [OIII] emissivity in the models we considered, with the assumed ionizing SED and gas metallicity. While these give a general description of the ionized gas, they cannot replace full photoionization models, where the emissivities can be different by a factor of $\sim$1.5. 

\begin{table}
	\centering 
	\tablewidth{0.8\linewidth} 
\begin{tabular}{|l|l|l|l|l|}
	\hline
 $\log U$ & -2.0 & -2.5 & -3.0 & -3.5 \\
 \hline
 $f(\log U)$ & 0.63 & 0.60 & 0.41 & 0.12 \\
 \hline
\end{tabular}
\caption{Correction coefficients for the [OIII] line emissivity expression given in equation \ref{eq:5_3}.}
\label{t:oiii_emis_dependence}
\end{table}

\section{Mass outflow rate and energetics}\label{s:mass_and_energy}

Having separated the stationary and outflowing gas components, we now turn to estimate the outflowing gas mass and its kinetic energy. The \emph{ionized} gas mass in the outflow can be estimated either from the broad H$\alpha$ luminosity, or from the broad [OIII] luminosity. The gas mass is given by (see e.g., \citealt{baron17b, fiore17}):
\begin{equation}\label{eq:2}
	M_{\mathrm{out}} = \frac{\mu m_{\mathrm{H}} L_{\mathrm{broad\, line}}} {\gamma_{\mathrm{line}} n_{e}},
\end{equation}
where $\mu$ is the mass per hydrogen atom, which we fix at 1.4, $m_{\mathrm{H}}$ is the hydrogen mass, $L_{\mathrm{broad\,line}}$ is the extinction-corrected broad line luminosity ($L_{\mathrm{H\alpha}}$ or $L_{\mathrm{[OIII]}}$), $n_{\mathrm{e}}$ is the electron density in the outflowing gas, and $\gamma_{\mathrm{line}}$ is the effective line emissivity ($\gamma_{\mathrm{H\alpha}}$ or $\gamma_{\mathrm{[OIII]}}$), which is given in section \ref{s:line_emis}. The [OIII] line emissivity depends on the gas metallicity, which is typically unknown. Lacking additional information, we assume solar metallicity.

We also estimate the mass outflow rate, the kinetic power of the wind, and the kinetic coupling efficiency. The mass outflow rate is given by $\dot{M}_{\mathrm{out}} = M_{\mathrm{out}}/t_{\mathrm{out}}$, where $t_{\mathrm{out}} = r_{\mathrm{out}}/v_{\mathrm{out}}$. The kinetic power in the wind is given by $\dot{E}_{\mathrm{kin,\, out}} = \frac{1}{2} \dot{M}_{\mathrm{out}} v_{\mathrm{out}}^2$. The kinetic coupling efficiency relates the wind kinetic power to the AGN bolometric luminosity: $\epsilon = \dot{E}_{\mathrm{kin,\, out}}/L_{\mathrm{bol}}$ (see e.g., \citealt{harrison18}). 

The above estimates require knowledge of the location of the wind, the electron density in the wind, and the effective outflow velocity. Since we use spatially-integrated (1D) spectroscopy and photometry, our estimates are based on luminosity-averaged properties, in particular the mass-weighted average properties of the wind. We describe in section \ref{s:wind_location} our estimate of the wind location, which is based on the method presented in Paper I. We then describe various methods to estimate the electron density in the wind in section \ref{s:wind_density}. Finally, in section \ref{s:wind_velocity} we discuss our definition of outflow velocity.

\subsection{Effective wind location}\label{s:wind_location}

\begin{figure}
	\center
\includegraphics[width=3.in]{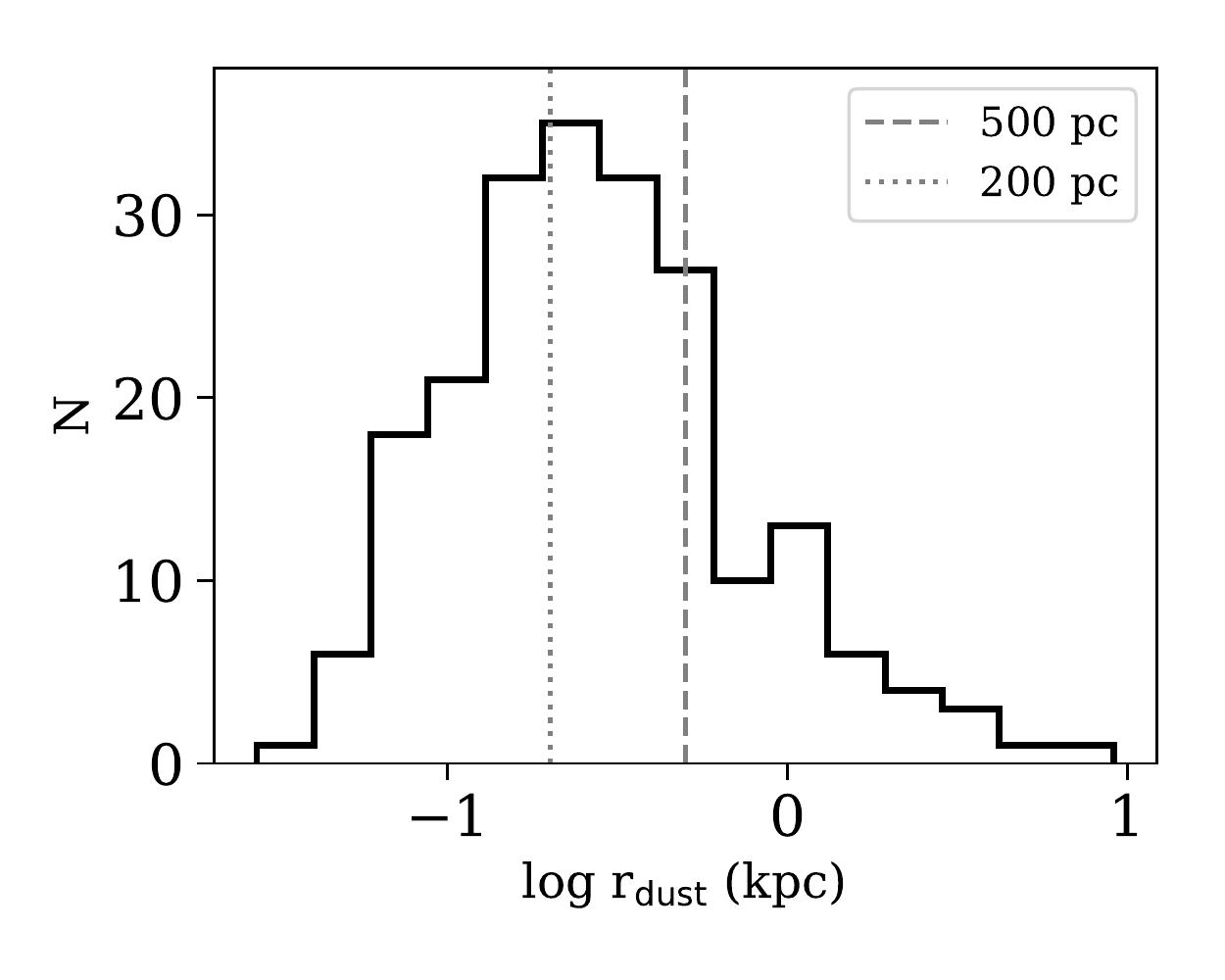}
\caption{The distribution of outflow locations in our sample. We refer the reader to Paper I for additional information about the dusty wind properties.}\label{f:rdust_distribution}
\end{figure}

In Paper I we argued that the outflowing gas in active galaxies contains dust, which is heated by the central AGN, and emits at mid-infrared wavelengths. We have shown that this emission component is detected in many type II AGN that host ionized gas outflows. This new component offers novel constrains on the outflow properties. Specifically, we used the dust temperature obtained from SED fitting and the AGN bolometric luminosity to estimate the distance of this dust from the central source. Our method provides luminosity-weighted, and therefore mass-weighted distance from the center of the galaxy. Since this dust is mixed with the outflowing gas, the method provides an estimate of the mass-weighted location of the outflow. In figure \ref{f:rdust_distribution} we show the distribution of outflow locations, traced by the dust infrared emission, for our sample. The distribution is similar to the distribution of our parent sample in Paper I. As discussed in Paper I, the uncertainty on the location is roughly 0.25 dex.

\subsection{Electron density}\label{s:wind_density}

The electron density of the outflowing gas is a major source of uncertainty. The most robust estimates are obtained from full photoionization modelling of the outflowing gas (e.g., \citealt{baron18, revalski18}). Such models require detailed, spatially-resolved spectroscopic observations, which are available only for a small number of objects, and hence they cannot be used to explore the entire range of population properties. Most earlier studies either assume a constant value for the electron density for all objects in the sample (e.g., \citealt{liu13b, harrison14, husemann16, fiore17}), or use the ratio of the weak [SII] emission lines to put limits on the density (e.g., \citealt{nesvadba06, nesvadba08, karouzos16b}). Typical [SII]-based estimates are in the range 200--1000 $\mathrm{cm^{-3}}$ (e.g, \citealt{fiore17}). Here we discuss the merits and the limitations of this and other methods used to determine the gas density in the outflow

\subsubsection{[SII] line ratio}\label{s:sii_density}

\begin{figure*}
	\center
\includegraphics[width=1.0\textwidth]{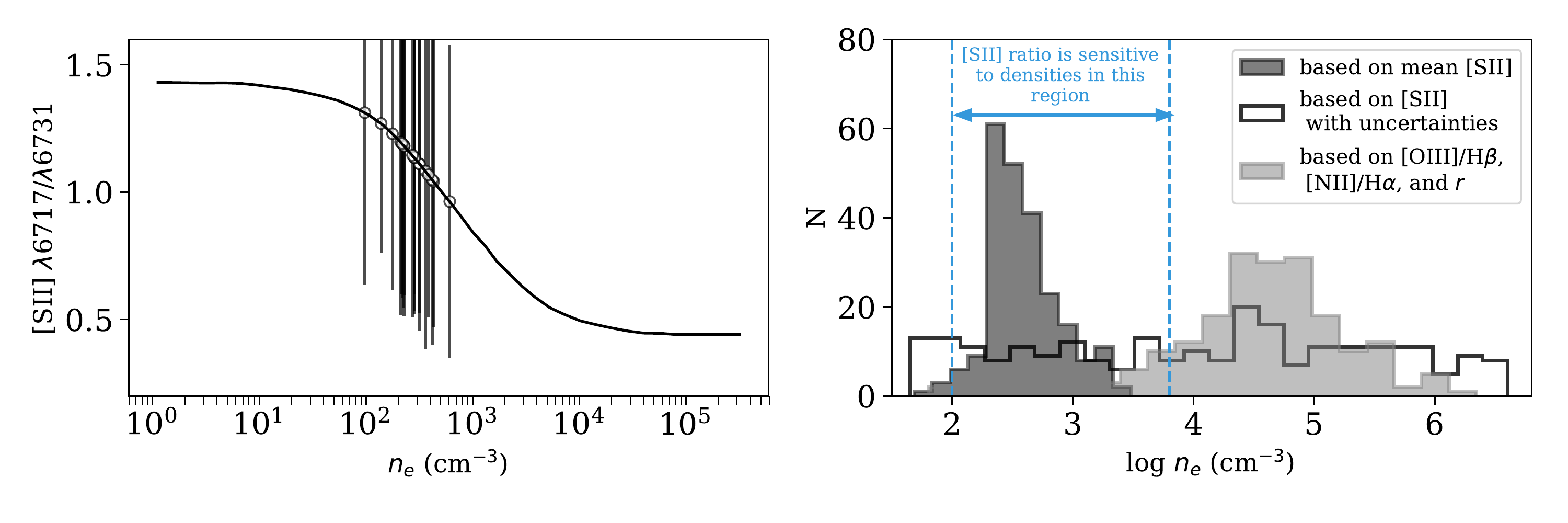}
\caption{ \textbf{Left panel:} [SII] $\lambda$6717/$\lambda$6731 doublet ratio versus electron density in ionized gas (black solid line). The errorbars represent 20 (out of 234) randomly-selected measurements for our sample, which are consistent, within their estimated errors, with the entire range of possible ratios. \textbf{Right panel:} the distribution of the electron density in the outflowing gas measured with two methods. The first is based on the commonly-used [SII] line ratio (dark grey histogram). Due to the significant uncertainties on these densities, we also show the distribution of densities after accounting for the measurement uncertainties (black empty histogram). The second method is based on optical line ratios and the known distance (light grey histogram; see text for details). The [SII]-based method is limited to a small range in densities (marked with a blue vertical lines). The second method covers the entire range of electron density. }\label{f:electron_density_distribution}
\end{figure*}

In the left panel of figure \ref{f:electron_density_distribution} we show the [SII] density diagnostic diagram. The [SII]-based estimator is sensitive to electron densities which are close to the critical density of the [SII]~$\lambda$ 6717,6731\AA\, lines: about 1600 and $1.5 \times 10^{4}\, \mathrm{cm^{-3}}$ for a gas temperature of $10^{4}$ K \citep{draine11}. The [SII] ratio saturates at the low ($n_{e} < 10^2\,\mathrm{cm^{-3}}$) and high ($n_{e} > 10^4\,\mathrm{cm^{-3}}$) density regimes, and cannot be used as a density estimator in these ranges. We use the broad [SII] lines measured in section \ref{s:spectral_props} to estimate the electron density in the outflow, and show 20 randomly-selected measurements on the diagram. Due to the significant uncertainty in the emission line decomposition, the ratios are consistent with the entire range of possible ratios, from roughly 0.5 to 1.5. In particular, most measurements are consistent with the high density limit of about $10^{4}\, \mathrm{cm^{-3}}$.

In the right panel of figure \ref{f:electron_density_distribution} we show two distributions of the electron density based on the [SII] lines. The first is using the initial guess provided to the fitting function and without taking account the large uncertainties. This distribution peaks at roughly $10^{2.5}$ $\mathrm{cm^{-3}}$, similar to the estimates by others (e.g. \citealt{fiore17}). Moreover, the exact location of the fit depends on the initial guess. We also show the distribution of [SII]-based densities after accounting for the measurement uncertainties. This, and the information provided in the left panel of the diagram, demonstrate that electron densities above about $10^{3.5}\,\mathrm{cm^{-3}}$ cannot be constrained using the [SII] line ratio.

The [SII]-based electron density is subjected to an additional systematic. For the range of ionization parameters considered here, the [OIII] and H$\alpha$ emission lines are emitted throughout most of the ionized cloud, while the [SII] lines are primarily emitted close to the ionization front in the cloud, where the electron density decreases with increasing depth into the cloud. Thus, the [OIII] and H$\alpha$ lines tend to trace higher electron density regions, compared to the [SII] lines. The difference between the electron density traced by [OIII] and H$\alpha$ and that traced by the [SII] becomes more significant for increasing ionization parameter, and for increasing hydrogen density in the cloud. Since the mass of the outflowing gas is estimated using the [OIII] or the H$\alpha$ emission line luminosities (equation \ref{eq:2}), the combination of [OIII] or H$\alpha$ line luminosity with [SII]-based electron density may result in inconsistent gas mass estimates, with significant deviations in cases with high ionization parameters and large hydrogen densities.

\subsubsection{[OIII]/H$\beta$, [NII]/H$\alpha$, and $r_{\mathrm{dust}}$}\label{s:ionization_par_density}

We now use the novel method presented in section \ref{s:elec_dens_est} to estimate the electron density in the outflowing gas. Given the coefficients in table \ref{t:ionization_parameter_coefficients}, we can use equations \ref{eq:5} and \ref{eq:4} to estimate the electron density, where we define the distance of the outflowing gas to be $r_{\mathrm{dust}}$, the mass-weighted location of the dust in the wind (see section \ref{s:wind_location}). We propagate the uncertainties of the line ratios, the effective location of the dust, and the analytical expression of the ionization parameter, to obtain the uncertainty on the electron density, which is roughly 0.6 dex. In the right panel of figure \ref{f:electron_density_distribution} we show the distribution of the electron densities in our sample using this method. The distribution is centered around $n_{\mathrm{e}} \sim 10^{4.5}\, \mathrm{cm^{-3}}$, suggesting that the [SII]-based method, shown on the left side of the diagram, underestimates the true electron density in the outflowing gas by roughly 2 orders of magnitude. 

\subsubsection{Auroral and transauroral [OII] and [SII] emission line ratios}\label{s:transauroral_density}

\citet{holt11} discussed the uncertainties involved in estimating the electron density using the [SII]~$\lambda$ 6717,6731\AA\, emission lines. They pointed out that this estimator is not sensitive to $n_{\mathrm{e}} \gtrsim 10^{4}\,\mathrm{cm^{-3}}$. To estimate the electron density, they used the transauroral and auroral emission lines [SII]~$\lambda$4068,4076\AA\, and [OII]~$\lambda$7318,7319,7330,7331\AA, which have critical densities around roughly $10^{6}\,\mathrm{cm^{-3}}$ (e.g., \citealt{draine11, holt11}). Their method is based on the \emph{summed} flux of each doublet, while the traditional [SII] method is based on the doublet flux ratio. As such, their estimated electron density does not suffer from the uncertainty involved in emission line decomposition, which is significant when using the traditional [SII] method. Their method is sensitive to a broad range of electron densities, from roughly $10^{2}\,\mathrm{cm^{-3}}$ to $10^{6}\,\mathrm{cm^{-3}}$. However, it is sensitive to reddening corrections.

\citet{rose18} applied this method to study the AGN-driven outflows in 9 ultra luminous infrared galaxies (ULIRGs) observed with VLT/Xshooter. The exceptional quality of the VLT/Xshooter observations allowed them to perform emission line decomposition of the weak emission lines [SII]~$\lambda$4068,4076\AA\, and [OII]~$\lambda$7318,7319,7330,7331\AA, and thus provide accurate estimates of the electron density in the outflowing gas. They found relatively high electron densities in the wind, from roughly $\log n_{e}(\mathrm{cm^{-3}}) = 3.4$ to $\log n_{e}(\mathrm{cm^{-3}}) = 4.8$  (see also \citealt{spence18}). \citet{santoro18} applied the method to an additional AGN, finding a remarkably high density of $\log n_{e}(\mathrm{cm^{-3}}) \simeq 5.5$. These estimates exceed the densities typically assumed or measured for ionized outflows in active galaxies. Our independent method confirms this in a much larger sample of non-ULIRG type II AGN. The densities we infer are somewhat larger than those found by \citet{rose18}. A detailed comparison between the samples should be carried out with caution, since the two have very different size and consist of different types of objects, i.e. low luminosity type II AGN versus ULIRGs.

The two methods are sensitive to a large range of electron densities, and are expected to give consistent results. There are several differences between the method presented in \citet{holt11} and our method. First, we assume that the distance of the gas from the central source is known, which is not required in the auroral and transauroral emission lines method. Furthermore, the method presented in \citet{holt11} can be used to put constrains on the dust reddening in the gas, while we need to estimate it using the observed H$\alpha$/H$\beta$ line ratio. However, the auroral and transauroral emission lines are very weak, and can only be detected in high-quality observations. In addition, this method requires a larger wavelength coverage, roughly from 3700\AA\, and 8000\AA. Therefore, while less general, our method is expected to be more practical in many cases.

\subsection{Velocity}\label{s:wind_velocity}

There are several uncertainties associated with the estimate of the bulk velocity of the outflow (e.g., \citealt{liu13b, harrison14, karouzos16b, fiore17, baron18}). First, since we work with spatially-integrated spectra, the wind geometry is unknown and we cannot correct for projection effects. Second, dust-extinction may affect the observed emission line profiles, resulting in underestimation of the outflow extent and kinematics. 

Various studies define the bulk velocity in different ways (e.g., \citealt{liu13b, karouzos16b, fiore17, harrison18}). We follow the definition by \citet{karouzos16b}, since the properties of their sample, such as wind velocity, AGN bolometric luminosity, and redshift, closely match the properties of our sample. According to Karouzos et al. (2016b), $v_{\mathrm{out}} = \sqrt{v_{\mathrm{shift}}^2 + \sigma^2 }$, where $v_{\mathrm{shift}}$ is the velocity shift of the broad emission line centroid with respect to the narrow lines (see figure \ref{f:broad_lines_kinematics}), and $\sigma$ is the velocity dispersion of the broad emission lines. According to Karouzos et al. (2016b), using different definitions can result in mass outflow rates and kinetic powers which differ from the one used here by a factor of 1--3.

\begin{figure}
	\center
\includegraphics[width=3.25in]{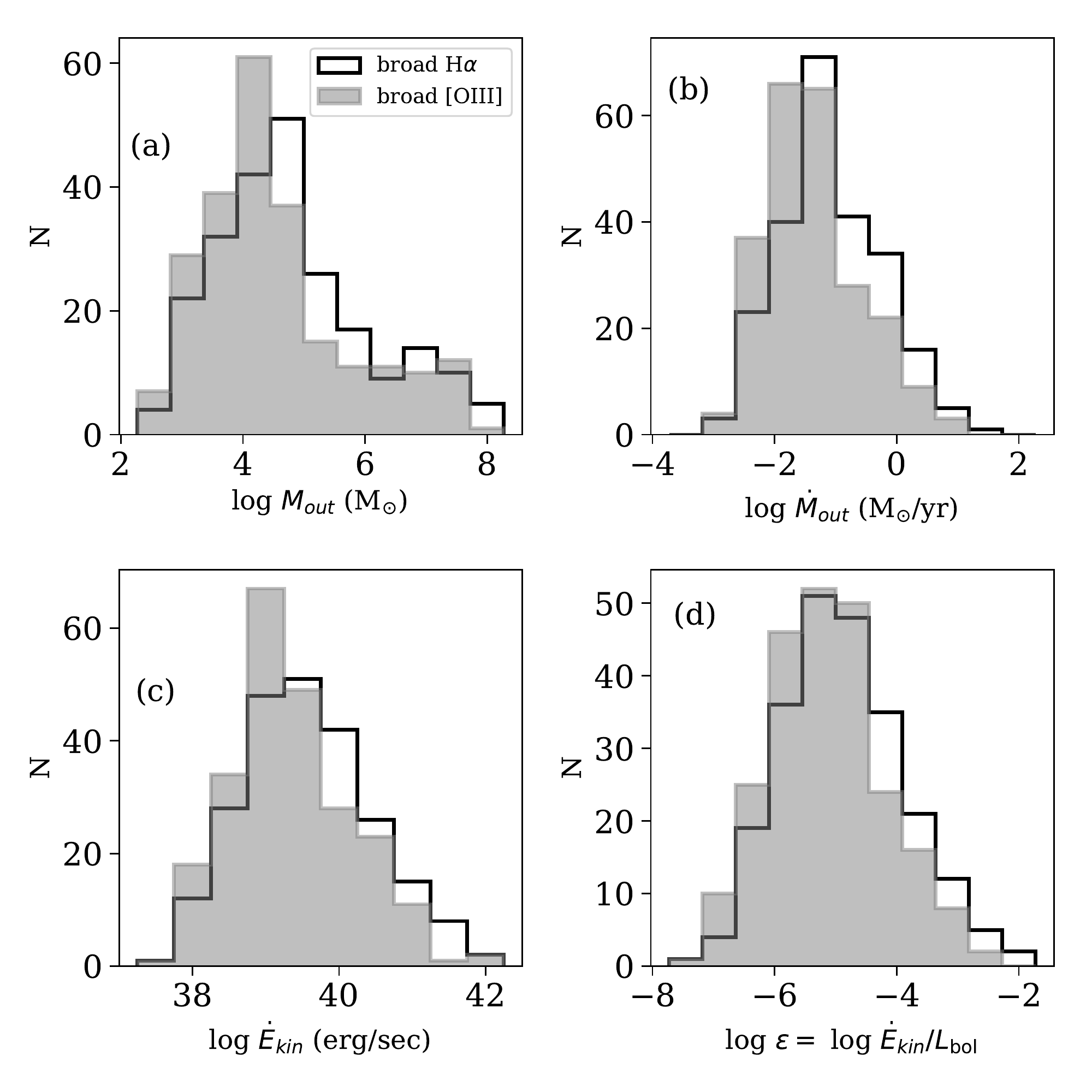}
\caption{The distributions of the ionized outflow properties in our sample. Panel (a) shows the ionized gas mass, panel (b) the mass outflow rate, panel (c) the kinetic power of the wind, and panel (d) the kinetic coupling efficiency. We show two estimates for these properties, one based on the dust-corrected broad H$\alpha$ luminosity (black histogram) and one on the dust-corrected broad [OIII] luminosity (grey histogram). The [OIII]-based estimates are consistent with the H$\alpha$-based estimates, suggesting that our assumption of solar metallicity is valid.}\label{f:mass_and_energetics}
\end{figure}

Having estimated the wind location and velocity, and the electron density in the outflow, we can now estimate the gas mass, mass outflow rate, and kinetic power of the wind. In figure \ref{f:mass_and_energetics} we show the distribution of ionized gas mass, mass outflow rate, kinetic power, and the kinetic coupling efficiency for the objects in our sample. We show two estimates for these properties, one based on the dust-corrected broad H$\alpha$ luminosity (black histogram), and one based on the dust-corrected broad [OIII] luminosity (grey histogram). The [OIII]-based estimates are consistent with the H$\alpha$-based estimates. This suggests that our assumption of solar metallicity is justified. 

Based on the broad H$\alpha$, we find that the median ionized gas mass in the outflow is roughly $M_{\mathrm{out}} \sim 10^{4}\, \mathrm{M_{\odot}}$, the median mass outflow rate is $\dot{M}_{\mathrm{out}} \sim 10^{-2}\, \mathrm{M_{\odot}/yr}$, and the median kinetic coupling efficiency is roughly $\epsilon = \dot{E}_{\mathrm{kin}}/L_{\mathrm{bol}} \sim 10^{-5}$. These values are 1--2 orders of magnitude lower than typical estimates in type II AGN with similar bolometric luminosity (see e.g., \citealt{karouzos16b, fiore17, harrison18}). These differences are the result of the much higher ionized gas density found in this work; roughly 2 orders of magnitude higher than in most previous studies. 

\section{Discussion}\label{s:disc}

In paper I of this series we have shown that the IR SED of the dusty wind component in a large number of type-II AGN can be used to place constraints on the wind location. Here, in paper II, we analysed a sub-sample of 234 type II AGN with signatures of ionized gas outflows in their optical spectra, and with a detection of a dusty wind component in mid-infrared wavelengths, with well determined temperature, covering factor, and location. The detection of the outflow in multiple emission lines (H$\beta$, [OIII], [OI], [NII], H$\alpha$, and [SII]) allowed us to accurately determine key-properties of the ionized flow. In particular, we introduced a novel method to estimate the ionization parameter of the ionized gas. The combination of known ionization parameter and location allow us to use a novel method to estimate the electron density in the ionized wind, and thus to estimate its mass outflow rate and kinetic power. 

Our analysis of spatially-integrated spectroscopy and photometry provides the mass-weighted average properties of the ionized outflows. These estimates are not as accurate as estimates that are based on spatially-resolved IFU observations, but they provide the largest compilation of ionized outflow properties to date, and can be used to study the statistical properties of winds in type II AGN. Below, we discuss our results in the broader context of AGN feedback (section \ref{s:agn_feedback}). We then elaborate on the neutral atomic gas which is associated with the observed ionized winds (section \ref{s:atomic_gas}).

\subsection{AGN feedback}\label{s:agn_feedback}

\begin{figure*}
\includegraphics[width=1\textwidth]{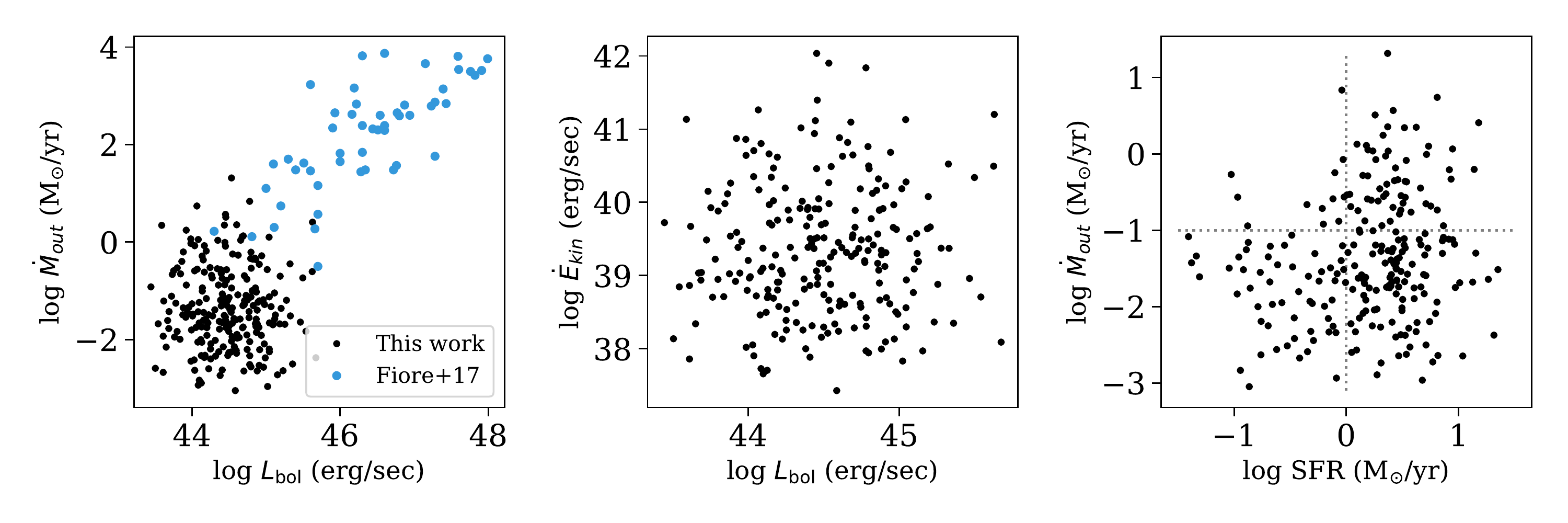}
\caption{\textbf{Left panel: }mass outflow rate versus AGN bolometric luminosity in our sample (black), compared to the measurements by \citet{fiore17} for ionized outflows (blue), Middle panel: the kinetic energy of the wind versus the AGN bolometric luminosity. \textbf{Right panel:} mass outflow rate versus star formation rate in the host galaxy. The uncertainty on the mass outflow rate and kinetic power is roughly 1 dex.}\label{f:mass_and_energetics_corrs}
\end{figure*}

AGN feedback, in the form of powerful outflows, is invoked as a way to couple the energy released by the accreting SMBH with the ISM of its host galaxy, thus providing a possible explanation for the observed correlations between the masses of SMBHs and their host bulges (e.g., \citealt{silk98,fabian99,king03,dimatteo05,springel05b}). A number of such models have successfully reproduced the observed correlations, by requiring that a significant amount of the accretion energy of the AGN will be mechanically-coupled to the ISM of the host galaxy ($\sim $5--10\% $L_{\mathrm{bol}}$; e.g., \citealt{fabian99,tremaine02, dimatteo05,springel05b,kurosawa09}). According to \citet{hopkins10}, in a two-stage feedback scenario, the initial energy requirement may be 10 times lower. 

The coupling efficiency used in hydrodynamic cosmological simulations is often defined as the fraction of the total energy carried out by the nuclear wind close to its launching location. It remains unclear how much of this energy is transferred to the galactic-scale winds, due to various energy losses such as shocks, radiation, and work that is done against the gravitational potential of the host galaxy (e.g., \citealt{veilleux17, richings18}; see \citealt{harrison18} for a comprehensive review). Furthermore, galactic-scale winds can be multi-phased, and it is often unclear what is the expected coupling efficiency of the different gas phases. Therefore, the coupling efficiencies required by numerical feedback models are not necessarily similar to kinetic coupling efficiencies derived from observations \citep{harrison18}. 

The typical mass outflow rate of the objects in our sample is roughly $10^{-2}\,\mathrm{M_{\odot}/yr}$, corresponding to kinetic coupling efficiencies around $\epsilon = \dot{E}_{\mathrm{kin}}/L_{\mathrm{bol}} \sim 10^{-5}$, 3--4 orders of magnitude lower than the typical requirement in most numerical simulations. Our estimates are about 1--2 orders of magnitude lower than typical estimates in AGN host galaxies (e.g., \citealt{harrison14,karouzos16b,fiore17,harrison18}), which are based on electron densities in the range $n_{e} \sim 10^{2} - 10^{3} \, \mathrm{cm^{-3}}$. Indeed, when we scale the mass outflow rates and coupling efficiencies found by \citet{fiore17} to have electron densities of $n_{e} \sim 10^{4.5} \, \mathrm{cm^{-3}}$, we find consistent results for AGN with the same bolometric luminosity. Furthermore, some of the earlier studies are based on IFU observations that target systems which show the most extreme outflow signatures. Our sample is less biased towards the most extreme outflow cases. 

In the left panel of figure \ref{f:mass_and_energetics_corrs} we show the mass outflow rate versus the AGN bolometric luminosity for the objects in our sample. For comparison, we show the measurements by \citet{fiore17} for ionized outflows which, for the same $L_{\mathrm{bol}}$, are much larger. Unlike \citet{fiore17}, we find no correlation between the mass outflow rate and the AGN bolometric luminosity in our sample. The middle panel shows the kinetic energy of the outflow versus $L_{\mathrm{bol}}$, where we find no correlation as well. In the right panel of figure \ref{f:mass_and_energetics_corrs} we show the mass outflow rate versus the star formation rate (SFR) in the host galaxy, where the latter is measured using the Dn4000 index (see details in Paper I). Surprisingly, the strongest ionized outflows in our sample are found in galaxies with high SFRs. This dependence might suggest that star formation is the main driver of the observed winds (see e.g., \citealt{wild10, cicone16}). Since both the narrow and broad kinematic components in all the objects in our sample are classified as AGN using line diagnostic diagrams, it is possible that the process that photoionizes the stationary (NLR) and outflowing gas is not necessarily the process that produces the observed outflows. While the AGN in our sample populate only 3 out of the 4 quadrants indicated in the diagram, which might suggest that there are no active galaxies with low SFRs and strong winds, the small correlation coefficient ($\rho = 0.16$) and the large uncertainty on the mass outflow rates, prevent us from drawing more definite conclusions. 

The lack of correlation between $\dot{M}_{\mathrm{out}}$ and $L_{\mathrm{bol}}$ might be related to the uncertainty in our mass outflow rate and bolometric luminosity measurements. Combining all the uncertainties discussed in section \ref{s:mass_and_energy}, we find the uncertainty on the mass outflow rate to be roughly 1 dex. The uncertainty on $L_{\mathrm{bol}}$ is 0.3--0.4 dex \citep{netzer09}. The combination of limited dynamical range (2 dex in $L_{\mathrm{bol}}$) and the above uncertainties is not enough to explain the \emph{complete} lack of correlation between $\dot{M}_{\mathrm{out}}$ and $L_{\mathrm{bol}}$. 

Finally, the lack of correlation with the AGN bolometric luminosity might be related to projection effects. If the AGN drives the observed winds through radiation pressure, we expect the gas to be accelerated within the AGN ionization cones, which are perpendicular to the line of sight in our type II AGN. Therefore, the measured wind velocity is underestimated by a factor of $\sin \alpha$, where $\alpha$ roughly represents the opening angle. Assuming that the scatter in the observed mass outflow rate is driven by scatter in the opening angle, such that $\log \dot{M}_{\mathrm{obs}} = \log \dot{M}_{\mathrm{outflow}} + \log(\sin \alpha)$, the scatter in the observed mass outflow rate is given by: $\Delta(\log \dot{M}_{\mathrm{obs}}) = \sqrt{ \Delta(\log \dot{M}_{\mathrm{outflow}})^2 + \Delta(\log(\sin \alpha))^2}$. Under these assumptions, the lack of correlation between $\dot{M}_{\mathrm{out}}$ and $L_{\mathrm{bol}}$ over 2 orders of magnitude implies that the scatter in $\sin \alpha$ is roughly 1-2 dex, depending on the exact shape of the $\sin \alpha$ distribution, and assuming $\log \frac{ \dot{M}_{\mathrm{out}}} {1 \, M_{\mathrm{\odot}}/\mathrm{yr}}  = 1.3 \log \frac{L_{\mathrm{bol}}}{10^{45}\,\mathrm{erg/sec}}$ \citep{fiore17}. We find this very large range in $\sin \alpha$ unlikely.

Given the above considerations, we suggest that the lack of correlation between $\dot{M}_{\mathrm{out}}$ and $L_{\mathrm{bol}}$ is not entirely due to measurement uncertainties and/or projection effects, and is partially driven by the scatter in the intrinsic properties of the winds. This is at odds with the recent results by \citet{fiore17}, who found such a correlation assuming a constant value electron density in all the objects in their sample. Our new estimates of the electron density for individual sources both decrease the measured mass outflow rates and introduce a scatter. The lack of intrinsic correlation can be attributed to several different factors. First, different outflow histories may result in different spatial extents of the observed winds. Furthermore, since outflows are most likely to go through the path of least resistance, the observed winds may have very different geometries, depending on the mass distribution in their host galaxy. 

\subsection{Neutral atomic gas in the outflow}\label{s:atomic_gas}

So far, we have focused on the warm ionized phase of the outflow, which is traced by strong optical emission lines. However, this gas can have large enough column density to be optically thick to the Lyman continuum radiation. For the ionization parameters found here, this requires a column in the range $N_{\mathrm{H}} = 10^{19.5}\,\mathrm{cm^{-2}}$ to $N_{\mathrm{H}} = 10^{20.5}\,\mathrm{cm^{-2}}$. The neutral gas at the back of such clouds is largely unconstrained by the observed optical lines, yet it may represent a significant fraction of the outflowing gas mass. The lower-ionization optical lines, such as [OI]~$\lambda \lambda$ 6300,6363\AA, [NII]~$\lambda \lambda$ 6548,6584\AA, and [SII]~$\lambda \lambda$ 6717,6731\AA, peak close to the ionization front. However, their emission drops significantly beyond it, and they cannot be used to place constraints on the fraction of neutral gas in the outflow. 

The NaID absorption line is commonly used to constrain neutral gas outflows in active galaxies (e.g., \citealt{rupke13, santoro18}). We examined the NaID absorption profile as a function of different outflow properties in our sample, and found no variation. However, the lack of variation does not necessarily imply that there is no neutral gas in the outflow, since the NaID absorption does not trace the same regions as the optical emission lines.

The dust mid-infrared emission can be used to constrain the neutral fraction of the outflows because the ratio of its infrared luminosity to the dust-corrected H$\alpha$ luminosity, L(IR)/L(H$\alpha$), is sensitive to the ionization parameter and to the total column density (ionized and neutral). Since the dust is mixed with the outflowing gas, L(IR)/L(H$\alpha$) should not depend on the gas covering factor. We find that the observed L(IR)/L(H$\alpha$) ratios in our sample are broadly consistent with those expected from photoionization models, given the empirically-estimated ionization parameters, for hydrogen column densities of $N_{\mathrm{H}} \sim 10^{20.5}-10^{21.5}\,\mathrm{cm^{-2}}$. This suggests significant amount of neutral atomic gas in most cases. It also supports the earlier conclusion (section \ref{s:photo_estimates} that most of the outflowing gas is radiation bounded. The inferred hydrogen column density implies that the mass of the neutral gas is a factor of a few larger than the observed ionized phase, with important implications to the mass and energetics of the winds.

Unfortunately, the dust infrared luminosity, L(IR), was estimated in Paper I though SED fitting with significant uncertainties of $\sim$ 0.5 dex. This, and the uncertainty on $U$ mentioned earlier, prevents us from reaching more definite conclusions about the exact amount of neutral atomic gas in individual cases. Furthermore, for a conical outflow, it is possible that the H$\alpha$ emission originating in the receding side of outflow is completely obscured by dust in the ISM of the host galaxy. While the ISM of the host galaxy is optically-thick to optical radiation, it is optically-thin to mid-infrared radiation, and the observed L(IR) traces the full extent of the dust in the outflow. Therefore, the measured L(IR)/L(H$\alpha$) ratio is probably overestimated by a factor of $\sim 2$.

\section{Summary and conclusions}\label{s:concs}

In the first paper of this series \citep{baron19}, we argued that the outflowing gas in active galaxies contains dust. The dust is heated by the central AGN and emits at mid-infrared wavelengths. We analysed the infrared spectral energy distribution of thousands of type II AGN and showed that this dust emission is detected in many systems that host ionized gas outflows. This infrared component offers novel constraints on the outflow properties, in particular, its location. In this work we focused on a subset of 234 galaxies from this sample, in which ionized outflows are detected in many optical emission lines ([OIII], H$\beta$, [OI], [NII], H$\alpha$, and [SII]), and there is a clear detection of a dusty wind component in mid-infrared wavelengths. The combination of the two allowed us to determine key properties of the gas in the wind, such as dust reddening, ionization state, and electron density, and thus constrain the mass and energetics of the winds. While our sample is based on spatially-integrated spectroscopy and photometry, and is thus less accurate than samples that are based on spatially-resolved observations, it forms the largest compilation of ionized outflow properties to date, and can be used to study the statistical properties of winds in type II AGN. Our results can be summarized as follows: 

\begin{itemize}

\item We presented a novel method to estimate the ionization parameter in the outflowing gas. We also provided more general expressions for the H$\alpha$ and [OIII] line emissivities, where the latter depends on the ionization state of the outflowing gas, and thus on the ionization parameter. The new line emissivities, which are then used to estimate the mass and energy of the ionized outflows, are expected to be more accurate than previously-used estimates.  

\item We presented a novel method to estimate the electron density in ionized gas, based on the strong optical line ratios [OIII]/H$\beta$ and [NII]/H$\alpha$, and on the known location of the gas. We applied this method to estimate the electron density in the observed outflows and found electron densities of $n_{\mathrm{e}} \sim 10^{4.5}\,\mathrm{cm^{-3}}$, which are about two orders of magnitude higher than most published estimates. We argue that the commonly-used method to estimate the electron density in the wind, which is based on the [SII] emission lines, underestimates the true densities by 1--2 orders of magnitude. Our estimates are more consistent with recent studies that are based on the density-sensitive auroral and transauroral [SII]~$\lambda$4068,4076\AA\, and [OII]~$\lambda$7318,7319,7330,7331\AA\, emission lines.

\item We found that the median \emph{ionized} gas mass in the outflow is roughly $M_{\mathrm{out}} \sim 10^{4}\, \mathrm{M_{\odot}}$, the median mass outflow rate is $\dot{M}_{\mathrm{out}} \sim 10^{-2}\, \mathrm{M_{\odot}/yr}$, and the median kinetic coupling efficiency is roughly $\epsilon = \dot{E}_{\mathrm{kin}}/L_{\mathrm{bol}} \sim 10^{-5}$. These values are 1--2 orders of magnitude lower than typical estimates in type II AGN with similar bolometric luminosity. This is a direct consequence of the high electron densities found in this study compared with the [SII]-based method. 

\item We found no correlation between the mass outflow rate of the ionized outflow and the AGN bolometric luminosity in our sample. This lack of correlation can be attributed to outflows having different locations, geometris, and different fractions of ionized gas in the outflowing clouds. Surprisingly, we found that the strongest outflows in our sample tend to occur in galaxies with high star formation rates, which might suggest that these are driven by supernovae in the host galaxy. Thus, the source that photoionizes the outflowing and stationary gas is not necessarily the source that drives the observed winds. 

\item Our study suggests the existence of a significant amount of neutral atomic gas at the back of the outflowing ionized gas clouds. We made an attempt to estimate the neutral gas fraction by combining the dust infrared luminosity with the H$\alpha$ luminosity. We suggest that the neutral gas mass at the back of the outflowing clouds is a factor of a few larger than the observed ionized gas mass, which has significant implications for the estimated mass and energetics of such flows. The large uncertainties on the dust infrared luminosity prevent us from estimating the neutral gas mass in individual sources.

\end{itemize}

\section*{Acknowledgments}
We thank the referee for useful comments and suggestions that helped improve this manuscript.
This work is supported in part by the Israel Science Foundation grant 284/13 and by DFG/DIP grant STE 1869/2-1 GE 625/17-1.
The spectroscopic analysis was made using IPython \citep{perez07}. We also used the following Python package: astropy\footnote{www.astropy.org/}.

This work made use of SDSS-III\footnote{www.sdss3.org} data. Funding for SDSS-III has been provided by the Alfred P. Sloan Foundation, the Participating Institutions, the National Science Foundation, and the U.S. Department of Energy Office of Science. SDSS-III is managed by the Astrophysical Research Consortium for the Participating Institutions of the SDSS-III Collaboration including the University of Arizona, the Brazilian Participation Group, Brookhaven National Laboratory, Carnegie Mellon University, University of Florida, the French Participation Group, the German Participation Group, Harvard University, the Instituto de Astrofisica de Canarias, the Michigan State/Notre Dame/JINA Participation Group, Johns Hopkins University, Lawrence Berkeley National Laboratory, Max Planck Institute for Astrophysics, Max Planck Institute for Extraterrestrial Physics, New Mexico State University, New York University, Ohio State University, Pennsylvania State University, University of Portsmouth, Princeton University, the Spanish Participation Group, University of Tokyo, University of Utah, Vanderbilt University, University of Virginia, University of Washington, and Yale University. 

\bibliographystyle{mn2e}
\bibliography{ref_outflows_AGN}

\begin{thebibliography}{77}
\expandafter\ifx\csname natexlab\endcsname\relax\def\natexlab#1{#1}\fi

\bibitem[{{Abazajian} {et~al}\mbox{.}(2009){Abazajian}, {Adelman-McCarthy},
  {Ag{\"u}eros}, {Allam}, {Allende Prieto}, {An}, {Anderson}, {Anderson},
  {Annis}, {Bahcall}, \& et~al.}]{abazajian09}
{Abazajian} K.~N. {et~al.}, 2009, \apjs, 182, 543

\bibitem[{{Baldwin}, {Phillips} \& {Terlevich}(1981){Baldwin}, {Phillips}, \&
  {Terlevich}}]{baldwin81}
{Baldwin} J.~A., {Phillips} M.~M., {Terlevich} R., 1981, \pasp, 93, 5

\bibitem[{{Baron} \& {Netzer}(2019)}]{baron19}
{Baron} D., {Netzer} H., 2019, \mnras, 482, 3915

\bibitem[{{Baron} {et~al}\mbox{.}(2017){Baron}, {Netzer}, {Poznanski},
  {Prochaska}, \& {F{\"o}rster Schreiber}}]{baron17b}
{Baron} D., {Netzer} H., {Poznanski} D., {Prochaska} J.~X., {F{\"o}rster
  Schreiber} N.~M., 2017, \mnras, 470, 1687

\bibitem[{{Baron} {et~al}\mbox{.}(2018){Baron}, {Netzer}, {Prochaska}, {Cai},
  {Cantalupo}, {Martin}, {Matuszewski}, {Moore}, {Morrissey}, \&
  {Neill}}]{baron18}
{Baron} D. {et~al.}, 2018, \mnras, 480, 3993

\bibitem[{{Binette}, {Wilson} \& {Storchi-Bergmann}(1996){Binette}, {Wilson},
  \& {Storchi-Bergmann}}]{binette96}
{Binette} L., {Wilson} A.~S., {Storchi-Bergmann} T., 1996, \aap, 312, 365

\bibitem[{{Brinchmann} {et~al}\mbox{.}(2004){Brinchmann}, {Charlot}, {White},
  {Tremonti}, {Kauffmann}, {Heckman}, \& {Brinkmann}}]{b04}
{Brinchmann} J., {Charlot} S., {White} S.~D.~M., {Tremonti} C., {Kauffmann} G.,
  {Heckman} T., {Brinkmann} J., 2004, \mnras, 351, 1151

\bibitem[{{Cano-D{\'{\i}}az} {et~al}\mbox{.}(2012){Cano-D{\'{\i}}az},
  {Maiolino}, {Marconi}, {Netzer}, {Shemmer}, \& {Cresci}}]{cano12}
{Cano-D{\'{\i}}az} M., {Maiolino} R., {Marconi} A., {Netzer} H., {Shemmer} O.,
  {Cresci} G., 2012, \aap, 537, L8

\bibitem[{{Cappellari}(2012)}]{cappellari12}
{Cappellari} M., 2012, {pPXF: Penalized Pixel-Fitting stellar kinematics
  extraction}. Astrophysics Source Code Library

\bibitem[{{Cappellari} \& {Emsellem}(2004)}]{cappellari04}
{Cappellari} M., {Emsellem} E., 2004, \pasp, 116, 138

\bibitem[{{Cardelli}, {Clayton} \& {Mathis}(1989){Cardelli}, {Clayton}, \&
  {Mathis}}]{cardelli89}
{Cardelli} J.~A., {Clayton} G.~C., {Mathis} J.~S., 1989, \apj, 345, 245

\bibitem[{{Cheung} {et~al}\mbox{.}(2016){Cheung}, {Bundy}, {Cappellari},
  {Peirani}, {Rujopakarn}, {Westfall}, {Yan}, {Bershady}, {Greene}, {Heckman},
  {Drory}, {Law}, {Masters}, {Thomas}, {Wake}, {Weijmans}, {Rubin}, {Belfiore},
  {Vulcani}, {Chen}, {Zhang}, {Gelfand}, {Bizyaev}, {Roman-Lopes}, \&
  {Schneider}}]{cheung16}
{Cheung} E. {et~al.}, 2016, \nat, 533, 504

\bibitem[{{Cicone}, {Maiolino} \& {Marconi}(2016){Cicone}, {Maiolino}, \&
  {Marconi}}]{cicone16}
{Cicone} C., {Maiolino} R., {Marconi} A., 2016, \aap, 588, A41

\bibitem[{{Cicone} {et~al}\mbox{.}(2014){Cicone}, {Maiolino}, {Sturm},
  {Graci{\'a}-Carpio}, {Feruglio}, {Neri}, {Aalto}, {Davies}, {Fiore},
  {Fischer}, {Garc{\'{\i}}a-Burillo}, {Gonz{\'a}lez-Alfonso},
  {Hailey-Dunsheath}, {Piconcelli}, \& {Veilleux}}]{cicone14}
{Cicone} C. {et~al.}, 2014, \aap, 562, A21

\bibitem[{{Cid Fernandes} {et~al}\mbox{.}(2010){Cid Fernandes},
  {Stasi{\'n}ska}, {Schlickmann}, {Mateus}, {Vale Asari}, {Schoenell}, \&
  {Sodr{\'e}}}]{cidfernandes10}
{Cid Fernandes} R., {Stasi{\'n}ska} G., {Schlickmann} M.~S., {Mateus} A., {Vale
  Asari} N., {Schoenell} W., {Sodr{\'e}} L., 2010, \mnras, 403, 1036

\bibitem[{{Di Matteo}, {Springel} \& {Hernquist}(2005){Di Matteo}, {Springel},
  \& {Hernquist}}]{dimatteo05}
{Di Matteo} T., {Springel} V., {Hernquist} L., 2005, \nat, 433, 604

\bibitem[{{Dopita} {et~al}\mbox{.}(2002){Dopita}, {Groves}, {Sutherland},
  {Binette}, \& {Cecil}}]{dopita02}
{Dopita} M.~A., {Groves} B.~A., {Sutherland} R.~S., {Binette} L., {Cecil} G.,
  2002, \apj, 572, 753

\bibitem[{{Draine}(2011)}]{draine11}
{Draine} B.~T., 2011, {Physics of the Interstellar and Intergalactic Medium}

\bibitem[{{Fabian}(1999)}]{fabian99}
{Fabian} A.~C., 1999, \mnras, 308, L39

\bibitem[{{Ferland} {et~al}\mbox{.}(2017){Ferland}, {Chatzikos}, {Guzm{\'a}n},
  {Lykins}, {van Hoof}, {Williams}, {Abel}, {Badnell}, {Keenan}, {Porter}, \&
  {Stancil}}]{ferland17}
{Ferland} G.~J. {et~al.}, 2017, Revista Mexicana de Astronomia y Astrofisica, 53, 385

\bibitem[{{Ferrarese} \& {Merritt}(2000)}]{ferrarese00}
{Ferrarese} L., {Merritt} D., 2000, \apjl, 539, L9

\bibitem[{{Fiore} {et~al}\mbox{.}(2017){Fiore}, {Feruglio}, {Shankar},
  {Bischetti}, {Bongiorno}, {Brusa}, {Carniani}, {Cicone}, {Duras}, {Lamastra},
  {Mainieri}, {Marconi}, {Menci}, {Maiolino}, {Piconcelli}, {Vietri}, \&
  {Zappacosta}}]{fiore17}
{Fiore} F. {et~al.}, 2017, \aap, 601, A143

\bibitem[{{Fischer} {et~al}\mbox{.}(2011){Fischer}, {Crenshaw}, {Kraemer},
  {Schmitt}, {Mushotsky}, \& {Dunn}}]{fischer11}
{Fischer} T.~C., {Crenshaw} D.~M., {Kraemer} S.~B., {Schmitt} H.~R.,
  {Mushotsky} R.~F., {Dunn} J.~P., 2011, \apj, 727, 71

\bibitem[{{Fischer} {et~al}\mbox{.}(2018){Fischer}, {Kraemer}, {Schmitt},
  {Longo Micchi}, {Crenshaw}, {Revalski}, {Vestergaard}, {Elvis}, {Gaskell},
  {Hamann}, {Ho}, {Hutchings}, {Mushotzky}, {Netzer}, {Storchi-Bergmann},
  {Straughn}, {Turner}, \& {Ward}}]{fischer18}
{Fischer} T.~C. {et~al.}, 2018, \apj, 856, 102

\bibitem[{{F{\"o}rster Schreiber} {et~al}\mbox{.}(2018){F{\"o}rster Schreiber},
  {{\"U}bler}, {Davies}, {Genzel}, {Wisnioski}, {Belli}, {Shimizu}, {Lutz},
  {Fossati}, {Herrera-Camus}, {Mendel}, {Tacconi}, {Wilman}, {Beifiori},
  {Brammer}, {Burkert}, {Carollo}, {Davies}, {Eisenhauer}, {Fabricius},
  {Lilly}, {Momcheva}, {Naab}, {Nelson}, {Price}, {Renzini}, {Saglia},
  {Sternberg}, {van Dokkum}, \& {Wuyts}}]{forster_schreiber18}
{F{\"o}rster Schreiber} N.~M. {et~al.}, 2018, ArXiv e-prints

\bibitem[{{Gebhardt} {et~al}\mbox{.}(2000){Gebhardt}, {Bender}, {Bower},
  {Dressler}, {Faber}, {Filippenko}, {Green}, {Grillmair}, {Ho}, {Kormendy},
  {Lauer}, {Magorrian}, {Pinkney}, {Richstone}, \& {Tremaine}}]{gebhardt00a}
{Gebhardt} K. {et~al.}, 2000, \apjl, 539, L13

\bibitem[{{Groves}, {Dopita} \& {Sutherland}(2004{\natexlab{a}}){Groves},
  {Dopita}, \& {Sutherland}}]{groves04a}
{Groves} B.~A., {Dopita} M.~A., {Sutherland} R.~S., 2004{\natexlab{a}}, \apjs,
  153, 9

\bibitem[{{Groves}, {Dopita} \& {Sutherland}(2004{\natexlab{b}}){Groves},
  {Dopita}, \& {Sutherland}}]{groves04b}
{Groves} B.~A., {Dopita} M.~A., {Sutherland} R.~S., 2004{\natexlab{b}}, \apjs,
  153, 75

\bibitem[{{G{\"u}ltekin} {et~al}\mbox{.}(2009){G{\"u}ltekin}, {Richstone},
  {Gebhardt}, {Lauer}, {Tremaine}, {Aller}, {Bender}, {Dressler}, {Faber},
  {Filippenko}, {Green}, {Ho}, {Kormendy}, {Magorrian}, {Pinkney}, \&
  {Siopis}}]{gultekin09}
{G{\"u}ltekin} K. {et~al.}, 2009, \apj, 698, 198

\bibitem[{{Harrison} {et~al}\mbox{.}(2014){Harrison}, {Alexander}, {Mullaney},
  \& {Swinbank}}]{harrison14}
{Harrison} C.~M., {Alexander} D.~M., {Mullaney} J.~R., {Swinbank} A.~M., 2014,
  \mnras, 441, 3306

\bibitem[{{Harrison} {et~al}\mbox{.}(2018){Harrison}, {Costa}, {Tadhunter},
  {Fl{\"u}tsch}, {Kakkad}, {Perna}, \& {Vietri}}]{harrison18}
{Harrison} C.~M., {Costa} T., {Tadhunter} C.~N., {Fl{\"u}tsch} A., {Kakkad} D.,
  {Perna} M., {Vietri} G., 2018, Nature Astronomy, 2, 198

\bibitem[{{Holt} {et~al}\mbox{.}(2011){Holt}, {Tadhunter}, {Morganti}, \&
  {Emonts}}]{holt11}
{Holt} J., {Tadhunter} C.~N., {Morganti} R., {Emonts} B.~H.~C., 2011, \mnras,
  410, 1527

\bibitem[{{Hopkins} \& {Elvis}(2010)}]{hopkins10}
{Hopkins} P.~F., {Elvis} M., 2010, \mnras, 401, 7

\bibitem[{{Husemann} {et~al}\mbox{.}(2016){Husemann}, {Scharw{\"a}chter},
  {Bennert}, {Mainieri}, {Woo}, \& {Kakkad}}]{husemann16}
{Husemann} B., {Scharw{\"a}chter} J., {Bennert} V.~N., {Mainieri} V., {Woo}
  J.-H., {Kakkad} D., 2016, \aap, 594, A44

\bibitem[{{Karouzos}, {Woo} \& {Bae}(2016{\natexlab{a}}){Karouzos}, {Woo}, \&
  {Bae}}]{karouzos16a}
{Karouzos} M., {Woo} J.-H., {Bae} H.-J., 2016{\natexlab{a}}, \apj, 819, 148

\bibitem[{{Karouzos}, {Woo} \& {Bae}(2016{\natexlab{b}}){Karouzos}, {Woo}, \&
  {Bae}}]{karouzos16b}
{Karouzos} M., {Woo} J.-H., {Bae} H.-J., 2016{\natexlab{b}}, \apj, 833, 171

\bibitem[{{Kauffmann} \& {Haehnelt}(2000)}]{kauffmann00}
{Kauffmann} G., {Haehnelt} M., 2000, \mnras, 311, 576

\bibitem[{{Kauffmann} {et~al}\mbox{.}(2003{\natexlab{a}}){Kauffmann},
  {Heckman}, {Tremonti}, {Brinchmann}, {Charlot}, {White}, {Ridgway},
  {Brinkmann}, {Fukugita}, {Hall}, {Ivezi{\'c}}, {Richards}, \&
  {Schneider}}]{kauff03a}
{Kauffmann} G. {et~al.}, 2003{\natexlab{a}}, \mnras, 346, 1055

\bibitem[{{Kauffmann} {et~al}\mbox{.}(2003{\natexlab{b}}){Kauffmann},
  {Heckman}, {White}, {Charlot}, {Tremonti}, {Brinchmann}, {Bruzual}, {Peng},
  {Seibert}, {Bernardi}, {Blanton}, {Brinkmann}, {Castander}, {Cs{\'a}bai},
  {Fukugita}, {Ivezic}, {Munn}, {Nichol}, {Padmanabhan}, {Thakar}, {Weinberg},
  \& {York}}]{kauff03b}
{Kauffmann} G. {et~al.}, 2003{\natexlab{b}}, \mnras, 341, 33

\bibitem[{{Kewley} {et~al}\mbox{.}(2013){Kewley}, {Dopita}, {Leitherer},
  {Dav{\'e}}, {Yuan}, {Allen}, {Groves}, \& {Sutherland}}]{kewley13}
{Kewley} L.~J., {Dopita} M.~A., {Leitherer} C., {Dav{\'e}} R., {Yuan} T.,
  {Allen} M., {Groves} B., {Sutherland} R., 2013, \apj, 774, 100

\bibitem[{{Kewley} {et~al}\mbox{.}(2001){Kewley}, {Dopita}, {Sutherland},
  {Heisler}, \& {Trevena}}]{kewley01}
{Kewley} L.~J., {Dopita} M.~A., {Sutherland} R.~S., {Heisler} C.~A., {Trevena}
  J., 2001, \apj, 556, 121

\bibitem[{{King}(2003)}]{king03}
{King} A., 2003, \apjl, 596, L27

\bibitem[{{Kurosawa}, {Proga} \& {Nagamine}(2009){Kurosawa}, {Proga}, \&
  {Nagamine}}]{kurosawa09}
{Kurosawa} R., {Proga} D., {Nagamine} K., 2009, \apj, 707, 823

\bibitem[{{Liu} {et~al}\mbox{.}(2013{\natexlab{a}}){Liu}, {Zakamska}, {Greene},
  {Nesvadba}, \& {Liu}}]{liu13a}
{Liu} G., {Zakamska} N.~L., {Greene} J.~E., {Nesvadba} N.~P.~H., {Liu} X.,
  2013{\natexlab{a}}, \mnras, 430, 2327

\bibitem[{{Liu} {et~al}\mbox{.}(2013{\natexlab{b}}){Liu}, {Zakamska}, {Greene},
  {Nesvadba}, \& {Liu}}]{liu13b}
{Liu} G., {Zakamska} N.~L., {Greene} J.~E., {Nesvadba} N.~P.~H., {Liu} X.,
  2013{\natexlab{b}}, \mnras, 436, 2576

\bibitem[{{Mullaney} {et~al}\mbox{.}(2013){Mullaney}, {Alexander}, {Fine},
  {Goulding}, {Harrison}, \& {Hickox}}]{mullaney13}
{Mullaney} J.~R., {Alexander} D.~M., {Fine} S., {Goulding} A.~D., {Harrison}
  C.~M., {Hickox} R.~C., 2013, \mnras, 433, 622

\bibitem[{{Nesvadba} {et~al}\mbox{.}(2008){Nesvadba}, {Lehnert}, {De Breuck},
  {Gilbert}, \& {van Breugel}}]{nesvadba08}
{Nesvadba} N.~P.~H., {Lehnert} M.~D., {De Breuck} C., {Gilbert} A.~M., {van
  Breugel} W., 2008, \aap, 491, 407

\bibitem[{{Nesvadba} {et~al}\mbox{.}(2006){Nesvadba}, {Lehnert}, {Eisenhauer},
  {Gilbert}, {Tecza}, \& {Abuter}}]{nesvadba06}
{Nesvadba} N.~P.~H., {Lehnert} M.~D., {Eisenhauer} F., {Gilbert} A., {Tecza}
  M., {Abuter} R., 2006, \apj, 650, 693

\bibitem[{{Netzer}(2009)}]{netzer09}
{Netzer} H., 2009, \mnras, 399, 1907

\bibitem[{{Netzer}(2013)}]{netzer13}
{Netzer} H., 2013, {The Physics and Evolution of Active Galactic Nuclei}

\bibitem[{P\'erez \& Granger(2007)}]{perez07}
P\'erez F., Granger B.~E., 2007, Computing in Science and Engineering, 9, 21

\bibitem[{{Perna} {et~al}\mbox{.}(2019){Perna}, {Cresci}, {Brusa}, {Lanzuisi},
  {Concas}, {Mainieri}, {Mannucci}, \& {Marconi}}]{perna19}
{Perna} M., {Cresci} G., {Brusa} M., {Lanzuisi} G., {Concas} A., {Mainieri} V.,
  {Mannucci} F., {Marconi} A., 2019, arXiv e-prints

\bibitem[{{Revalski} {et~al}\mbox{.}(2018){Revalski}, {Dashtamirova},
  {Crenshaw}, {Kraemer}, {Fischer}, {Schmitt}, {Gnilka}, {Schmidt}, {Elvis},
  {Fabbiano}, {Storchi-Bergmann}, {Maksym}, \& {Gandhi}}]{revalski18}
{Revalski} M. {et~al.}, 2018, ArXiv e-prints

\bibitem[{{Richings} \& {Faucher-Gigu{\`e}re}(2018)}]{richings18}
{Richings} A.~J., {Faucher-Gigu{\`e}re} C.-A., 2018, \mnras, 478, 3100

\bibitem[{{Rose} {et~al}\mbox{.}(2018){Rose}, {Tadhunter}, {Ramos Almeida},
  {Rodr{\'{\i}}guez Zaur{\'{\i}}n}, {Santoro}, \& {Spence}}]{rose18}
{Rose} M., {Tadhunter} C., {Ramos Almeida} C., {Rodr{\'{\i}}guez Zaur{\'{\i}}n}
  J., {Santoro} F., {Spence} R., 2018, \mnras, 474, 128

\bibitem[{{Rupke}, {G{\"u}ltekin} \& {Veilleux}(2017){Rupke}, {G{\"u}ltekin},
  \& {Veilleux}}]{rupke17}
{Rupke} D.~S.~N., {G{\"u}ltekin} K., {Veilleux} S., 2017, \apj, 850, 40

\bibitem[{{Rupke} \& {Veilleux}(2013)}]{rupke13}
{Rupke} D.~S.~N., {Veilleux} S., 2013, \apj, 768, 75

\bibitem[{{Salim} {et~al}\mbox{.}(2007){Salim}, {Rich}, {Charlot},
  {Brinchmann}, {Johnson}, {Schiminovich}, {Seibert}, {Mallery}, {Heckman},
  {Forster}, {Friedman}, {Martin}, {Morrissey}, {Neff}, {Small}, {Wyder},
  {Bianchi}, {Donas}, {Lee}, {Madore}, {Milliard}, {Szalay}, {Welsh}, \&
  {Yi}}]{salim07}
{Salim} S. {et~al.}, 2007, \apjs, 173, 267

\bibitem[{{Santoro} {et~al}\mbox{.}(2018){Santoro}, {Rose}, {Morganti},
  {Tadhunter}, {Oosterloo}, \& {Holt}}]{santoro18}
{Santoro} F., {Rose} M., {Morganti} R., {Tadhunter} C., {Oosterloo} T.~A.,
  {Holt} J., 2018, \aap, 617, A139

\bibitem[{{Sharp} \& {Bland-Hawthorn}(2010)}]{sharp10}
{Sharp} R.~G., {Bland-Hawthorn} J., 2010, \apj, 711, 818

\bibitem[{{Silk} \& {Rees}(1998)}]{silk98}
{Silk} J., {Rees} M.~J., 1998, \aap, 331, L1

\bibitem[{{Skrutskie} {et~al}\mbox{.}(2006){Skrutskie}, {Cutri}, {Stiening},
  {Weinberg}, {Schneider}, {Carpenter}, {Beichman}, {Capps}, {Chester},
  {Elias}, {Huchra}, {Liebert}, {Lonsdale}, {Monet}, {Price}, {Seitzer},
  {Jarrett}, {Kirkpatrick}, {Gizis}, {Howard}, {Evans}, {Fowler}, {Fullmer},
  {Hurt}, {Light}, {Kopan}, {Marsh}, {McCallon}, {Tam}, {Van Dyk}, \&
  {Wheelock}}]{skrutskie06}
{Skrutskie} M.~F. {et~al.}, 2006, \aj, 131, 1163

\bibitem[{{Spence} {et~al}\mbox{.}(2018){Spence}, {Tadhunter}, {Rose}, \&
  {Rodr{\'{\i}}guez Zaur{\'{\i}}n}}]{spence18}
{Spence} R.~A.~W., {Tadhunter} C.~N., {Rose} M., {Rodr{\'{\i}}guez
  Zaur{\'{\i}}n} J., 2018, \mnras, 478, 2438

\bibitem[{{Springel}, {Di Matteo} \& {Hernquist}(2005){Springel}, {Di Matteo},
  \& {Hernquist}}]{springel05b}
{Springel} V., {Di Matteo} T., {Hernquist} L., 2005, \mnras, 361, 776

\bibitem[{{Stern}, {Laor} \& {Baskin}(2013){Stern}, {Laor}, \&
  {Baskin}}]{stern13}
{Stern} J., {Laor} A., {Baskin} A., 2013, \mnras, 438, 901

\bibitem[{{Tadhunter} {et~al}\mbox{.}(2018){Tadhunter}, {Rodr{\'{\i}}guez
  Zaur{\'{\i}}n}, {Rose}, {Spence}, {Batcheldor}, {Berg}, {Ramos Almeida},
  {Spoon}, {Sparks}, \& {Chiaberge}}]{tadhunter18}
{Tadhunter} C. {et~al.}, 2018, \mnras, 478, 1558

\bibitem[{{Tremaine} {et~al}\mbox{.}(2002){Tremaine}, {Gebhardt}, {Bender},
  {Bower}, {Dressler}, {Faber}, {Filippenko}, {Green}, {Grillmair}, {Ho},
  {Kormendy}, {Lauer}, {Magorrian}, {Pinkney}, \& {Richstone}}]{tremaine02}
{Tremaine} S. {et~al.}, 2002, \apj, 574, 740

\bibitem[{{Tremonti} {et~al}\mbox{.}(2004){Tremonti}, {Heckman}, {Kauffmann},
  {Brinchmann}, {Charlot}, {White}, {Seibert}, {Peng}, {Schlegel}, {Uomoto},
  {Fukugita}, \& {Brinkmann}}]{t04}
{Tremonti} C.~A. {et~al.}, 2004, \apj, 613, 898

\bibitem[{{Vazdekis} {et~al}\mbox{.}(2010){Vazdekis},
  {S{\'a}nchez-Bl{\'a}zquez}, {Falc{\'o}n-Barroso}, {Cenarro}, {Beasley},
  {Cardiel}, {Gorgas}, \& {Peletier}}]{vazdekis10}
{Vazdekis} A., {S{\'a}nchez-Bl{\'a}zquez} P., {Falc{\'o}n-Barroso} J.,
  {Cenarro} A.~J., {Beasley} M.~A., {Cardiel} N., {Gorgas} J., {Peletier}
  R.~F., 2010, \mnras, 404, 1639

\bibitem[{{Veilleux} {et~al}\mbox{.}(2017){Veilleux}, {Bolatto}, {Tombesi},
  {Mel{\'e}ndez}, {Sturm}, {Gonz{\'a}lez-Alfonso}, {Fischer}, \&
  {Rupke}}]{veilleux17}
{Veilleux} S., {Bolatto} A., {Tombesi} F., {Mel{\'e}ndez} M., {Sturm} E.,
  {Gonz{\'a}lez-Alfonso} E., {Fischer} J., {Rupke} D.~S.~N., 2017, \apj, 843,
  18

\bibitem[{{Veilleux} {et~al}\mbox{.}(2013){Veilleux}, {Mel{\'e}ndez}, {Sturm},
  {Gracia-Carpio}, {Fischer}, {Gonz{\'a}lez-Alfonso}, {Contursi}, {Lutz},
  {Poglitsch}, {Davies}, {Genzel}, {Tacconi}, {de Jong}, {Sternberg}, {Netzer},
  {Hailey-Dunsheath}, {Verma}, {Rupke}, {Maiolino}, {Teng}, \&
  {Polisensky}}]{veilleux13}
{Veilleux} S. {et~al.}, 2013, \apj, 776, 27

\bibitem[{{Veilleux} \& {Osterbrock}(1987)}]{veilleux87}
{Veilleux} S., {Osterbrock} D.~E., 1987, \apjs, 63, 295

\bibitem[{{Villar-Mart{\'{\i}}n} {et~al}\mbox{.}(2016){Villar-Mart{\'{\i}}n},
  {Arribas}, {Emonts}, {Humphrey}, {Tadhunter}, {Bessiere}, {Cabrera Lavers},
  \& {Ramos Almeida}}]{villar_martin16}
{Villar-Mart{\'{\i}}n} M., {Arribas} S., {Emonts} B., {Humphrey} A.,
  {Tadhunter} C., {Bessiere} P., {Cabrera Lavers} A., {Ramos Almeida} C., 2016,
  \mnras, 460, 130

\bibitem[{{Wild}, {Heckman} \& {Charlot}(2010){Wild}, {Heckman}, \&
  {Charlot}}]{wild10}
{Wild} V., {Heckman} T., {Charlot} S., 2010, \mnras, 405, 933

\bibitem[{{Wright} {et~al}\mbox{.}(2010){Wright}, {Eisenhardt}, {Mainzer},
  {Ressler}, {Cutri}, {Jarrett}, {Kirkpatrick}, {Padgett}, {McMillan},
  {Skrutskie}, {Stanford}, {Cohen}, {Walker}, {Mather}, {Leisawitz}, {Gautier},
  {McLean}, {Benford}, {Lonsdale}, {Blain}, {Mendez}, {Irace}, {Duval}, {Liu},
  {Royer}, {Heinrichsen}, {Howard}, {Shannon}, {Kendall}, {Walsh}, {Larsen},
  {Cardon}, {Schick}, {Schwalm}, {Abid}, {Fabinsky}, {Naes}, \&
  {Tsai}}]{wright10}
{Wright} E.~L. {et~al.}, 2010, \aj, 140, 1868

\bibitem[{{Zakamska} {et~al}\mbox{.}(2016){Zakamska}, {Hamann}, {P{\^a}ris},
  {Brandt}, {Greene}, {Strauss}, {Villforth}, {Wylezalek}, {Alexandroff}, \&
  {Ross}}]{zakamska16}
{Zakamska} N.~L. {et~al.}, 2016, \mnras

\bibitem[{{Zubovas} \& {Nayakshin}(2014)}]{zubovas14}
{Zubovas} K., {Nayakshin} S., 2014, \mnras, 440, 2625

\end{thebibliography}

\clearpage

\appendix

\section{Photoionization models}\label{a:photoionization_models}

We model the central source using standard assumptions about AGN SEDs. The SED consists of a combination of an optical-UV continuum emitted by an optically-thick geometrically-thin accretion disk, and an additional X-ray power-law source that extends to 50 keV with a photon index of $\Gamma = 1.9$. The normalisation of the UV (2500\AA) to X-ray (2 keV) flux is defined by $\alpha_{OX}$, which we take to be 1.37.

\begin{table}
	\centering 
	\tablewidth{0.8\linewidth} 
\begin{tabular}{|l|l|l|l|l|}
	\hline
 Model & $\log M_{\mathrm{BH}}\, (M_{\odot})$ & $L/L_{\mathrm{Edd}}$ & $a$ & $<h\nu>_{\mathrm{ion}}$ (Ryd)\\
 \hline
 1 & 8 & 0.12 & 0.7 & 3.15 \\
 2 & 7 & 0.14 & 0 & 2.56 \\
 3 & 9 & 0.15 & 0.998 & 2.65 \\
 4 & 7 & 0.36 & 0.7 & 4.17 \\
 \hline
\end{tabular}
\caption{Properties of the SEDs we consider in the modelling. Column (2): BH mass, column (3): Eddington ratio, column (4): BH spin, and column (5): mean energy of an ionizing photon.}
\label{t:SEDs}
\end{table}

The grid of models consists of a geometrically-thin, optically-thick, shells of dusty gas, with ISM-type grains and a density of $n_{\mathrm{H}} = 10^{4}\, \mathrm{cm^{-3}}$. We used a range of gas metallicities: 0.5, 0.65, 0.8, 1, 1.5, and 2$\mathrm{Z_{\odot}}$, and ionization parameter in the range $\log U = -3.8$ to $\log U = -2$. We consider four different shapes for the ionizing continuum SED, and we list their properties in table \ref{t:SEDs}. Specifically, the SEDs have different slopes, with mean energies of ionizing photon of 2.56, 2.65, 3.15, and 4.17 Ryd. The ranges of metallicities, ionization parameters, and ionizing SEDs were chosen to fully cover the range of possible gas and AGN properties, and thus to provide a general description of the sources in our sample. For the typical AGN in our sample, we use SED 2 (mean energy of an ionizing photon of 2.56 Ryd), where the number of ionizing photons is given by:
\begin{equation}\label{eq:7}
	{\log Q(\mathrm{Lyman}) = 10.06 + \log L_{\mathrm{bol}}},
\end{equation}
where $L_{\mathrm{bol}}$ is the AGN bolometric luminosity, which can be estimated using narrow emission line luminosities (see \citealt{netzer09}). Using the other SEDs (models 1, 3, and 4) result in a value which is different by roughly 10\% from the value given in equation \ref{eq:7}. Substituting equation \ref{eq:7} into equation \ref{eq:3} results in the expression given in equation \ref{eq:4} in the main text.

\begin{figure}
	\center
\includegraphics[width=3.25in]{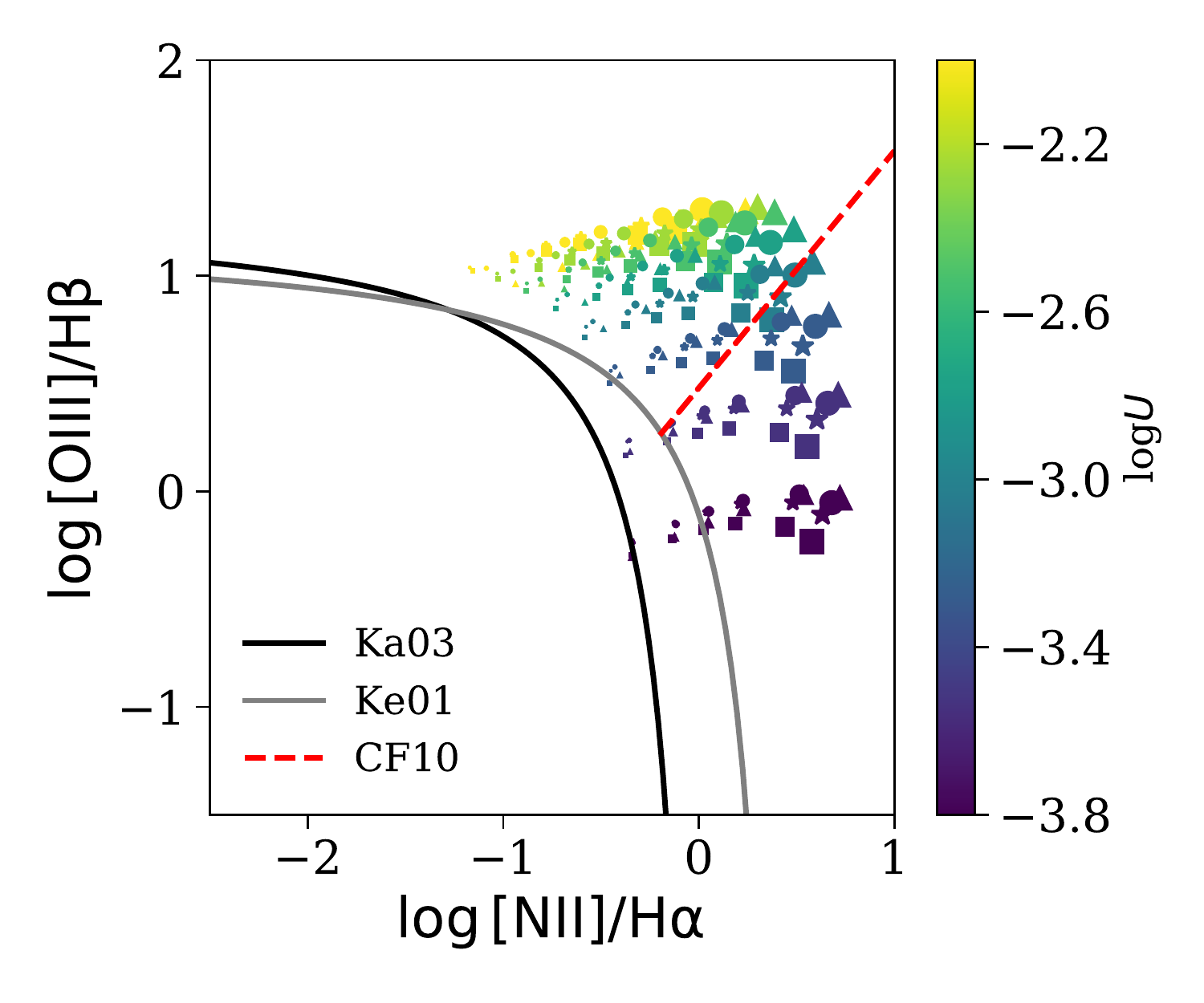}
\caption{Predicted [OIII]/H$\beta$ versus [NII]/H$\alpha$ for the 192 models considered here. The colour of the markers represents different ionization parameters, ranging from $\log U = -3.8$ to $\log U = -2$. The size of the markers represent the metallicity, where the smallest markers correspond to 0.5$\mathrm{Z_{\odot}}$, and the largest markers to 2$\mathrm{Z_{\odot}}$. The shapes of the markers represent the ionizing SED, where the squares, stars, circles, and triangles correspond to SEDs with a mean energy of ionizing photon of 2.56, 2.65, 3.15, and 4.17 Ryd respectively. }\label{f:all_models_BPT_diagram}
\end{figure}

We run a grid of 192 models with version 17.00 of \cloudy\ \citep{ferland17}. In figure \ref{f:all_models_BPT_diagram} we show the predicted [OIII]/H$\beta$ versus [NII]/H$\alpha$ by the different models. The colour represents the ionization parameter of the gas, and the size of the markers represents the metallicity, where the smallest markers correspond to 0.5$\mathrm{Z_{\odot}}$, and the largest markers to 2$\mathrm{Z_{\odot}}$. The shapes of the markers represent the ionizing SED, where the squares, stars, circles, and triangles correspond to SEDs with a mean energy of ionizing photon of 2.56, 2.65, 3.15, and 4.17 Ryd respectively. The predictions of these models are in agreement with predictions by other studies (e.g., \citealt{groves04a, groves04b}).

The models presented in figure \ref{f:all_models_BPT_diagram} are calculated under the assumption of constant density in the dusty gas. We also examined a similar grid of models, where we assume a constant total pressure (gas and radiation) in the gas. We find that the two sets of models give consistent optical line ratios for a given ionization parameter, and thus our conclusions do not depend on the assumption of constant density versus constant pressure in the gas. This is expected since we model gas clouds with ionization parameters of $\log U \leq -2$, where gas pressure is larger than radiation pressure. Therefore, these constant-pressure models do not represent a case of radiation pressure confinement (RPC), and are effectively constant-density models. RPC takes place when the radiation pressure is significantly larger than the gas pressure \citep{dopita02, stern13}, which is achieved for gas clouds with $\log U \geq -2$. The calculations of \citet{stern13} show that such models can reproduce the spectrum of Seyfert galaxies with the most extreme [OIII]/H$\beta$ line ratios (e.g., figure 10 in \citealt{stern13}), but cannot describe the bulk of the population. In particular, they do not cover a large region of the BPT diagram where many of our sources reside (figure 2). In such models, a range of ionization parameters (from $\log U \sim 3$ to $\log U \sim -1$) produce similar optical line ratios, and thus the line ratios cannot be used to place constrains on the average ionization parameter in the gas. In the rest of the section, we examine only constant-density models, since constant-pressure models give consistent results for our selected gas properties. 

Figure \ref{f:all_models_BPT_diagram} shows that the metallicity and the shape of the ionizing continuum are degenerate. That is, different combinations of gas metallicity and SED shape result in similar [OIII]/H$\beta$ and [NII]/H$\alpha$ line ratios. Therefore, these two properties cannot be determined from the optical line ratios alone, and require additional observables. On the other hand, the colour of the markers varies smoothly across the diagram, suggesting that the ionization parameter can be determined from the [OIII]/H$\beta$ and [NII]/H$\alpha$ line ratios, almost regardless of the gas metallicity or the shape of the ionizing continuum. 

\begin{figure}
	\center
\includegraphics[width=3.25in]{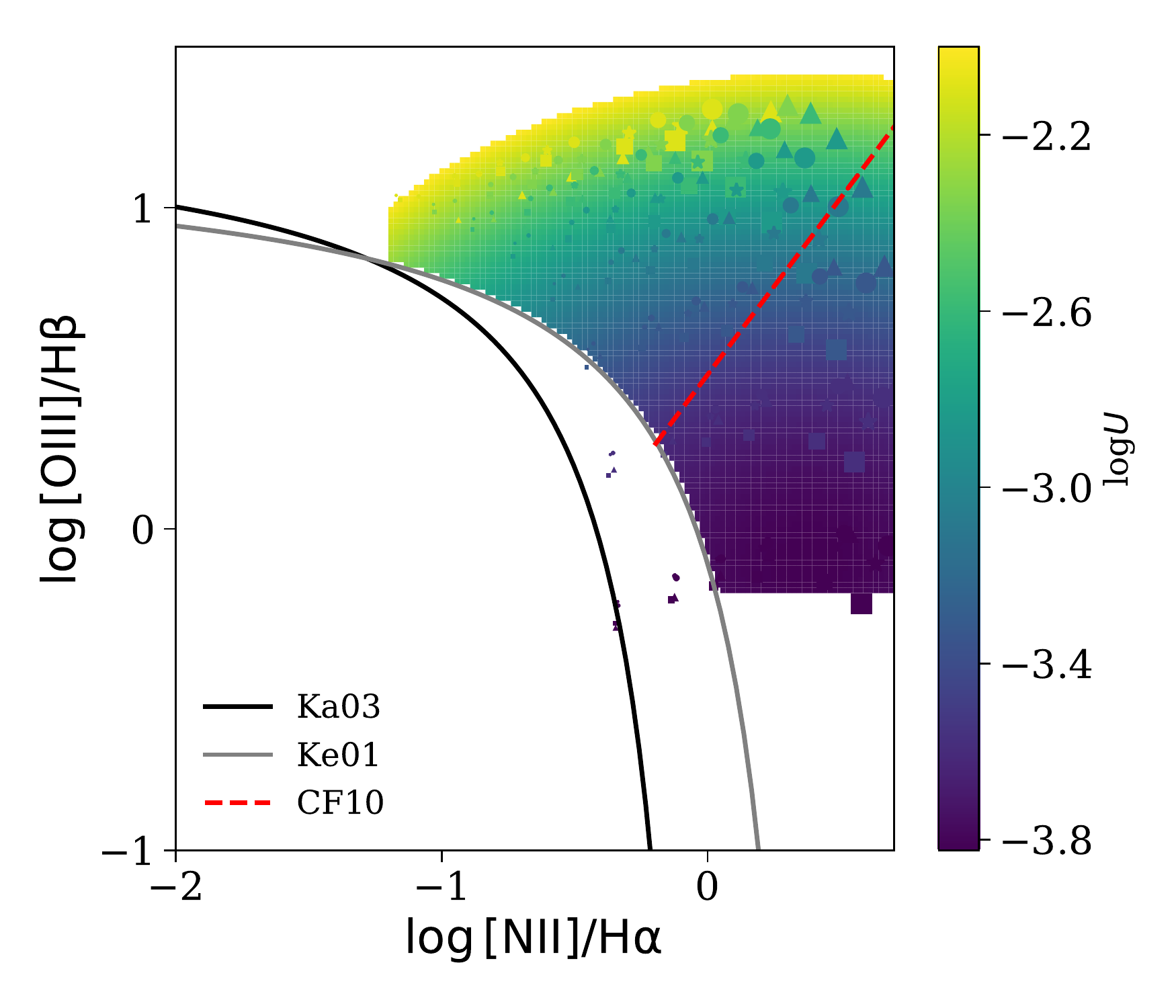}
\caption{Similar to figure \ref{f:all_models_BPT_diagram} but combined with a continuous range of $U$ obtained from equation \ref{eq:5}, which is marked here by continuous background colours (see scale on the right hand side of the diagram). The similarity in colour between the points and the background suggests that equation \ref{eq:5} gives a good fit to the ionization parameter. }\label{f:photoionization_models_and_analytical_formula}
\end{figure}

\begin{figure}
	\center
\includegraphics[width=3.25in]{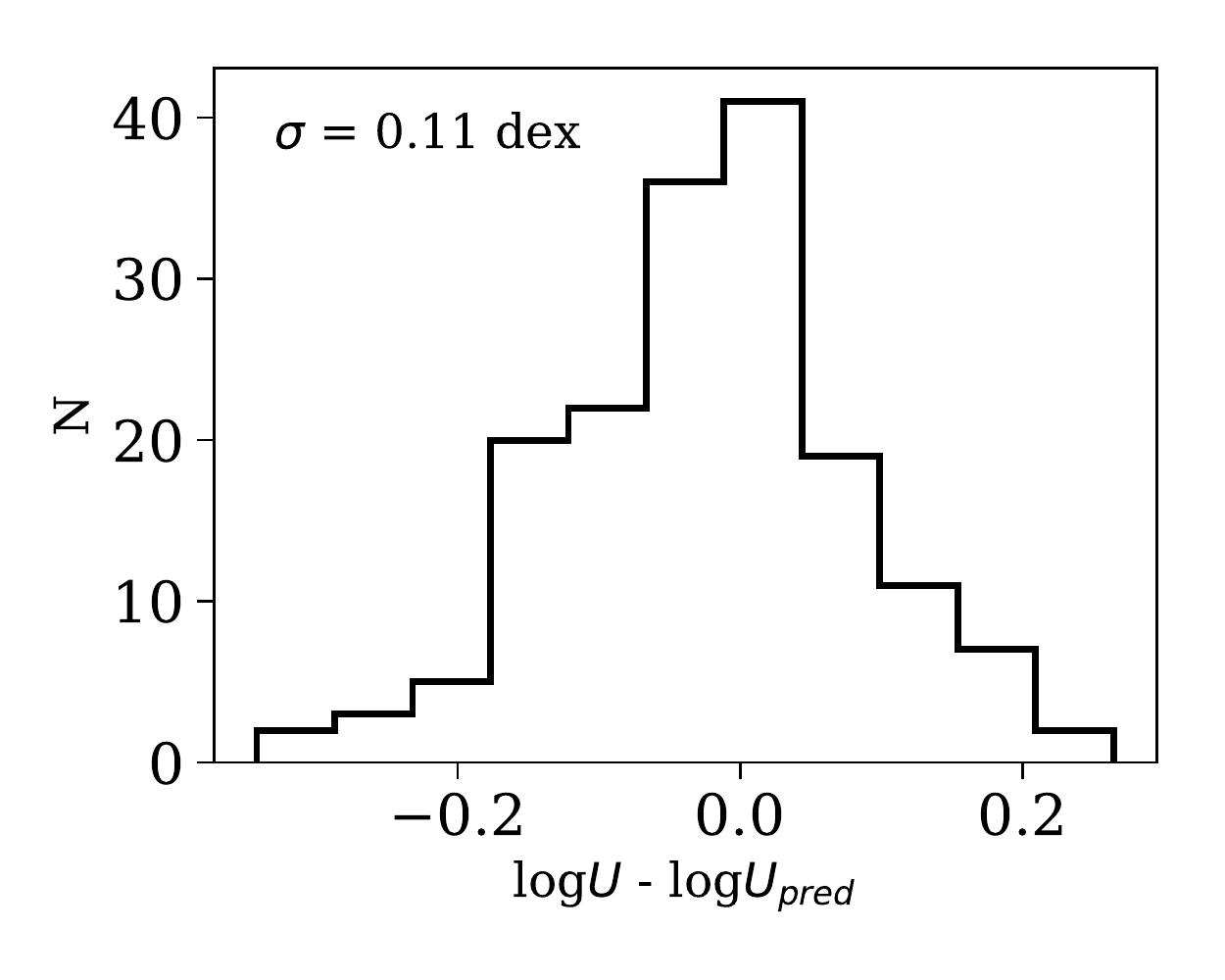}
\caption{The distribution of the residuals between the ionization parameter predicted by the specific photoionization models and the ionization parameter derived using equation \ref{eq:5}. The standard deviation is $\sigma = 0.11$ dex.}\label{f:photoionization_model_residuals}
\end{figure}

The ionization parameter expression derived from our modeling is shown in equation \ref{eq:5}. We find the coefficients $a_{1}$, $a_{2}$, $a_{3}$, $a_{4}$, and $a_{5}$, that minimize the residuals between the ionization parameter predicted by the photoionization models and the ionization parameter derived using equation \ref{eq:5}. We list the best-fitting coefficients in table \ref{t:ionization_parameter_coefficients}. In figure \ref{f:photoionization_models_and_analytical_formula} we show the predicted [OIII]/H$\beta$ versus [NII]/H$\alpha$ by the different models (same as figure \ref{f:all_models_BPT_diagram}), where the background colour represents the ionization parameter that is derived according to equation \ref{eq:5}. Overall, there is a good agreement between the ionization parameters predicted by the models and those derived using equation \ref{eq:5}. In figure \ref{f:photoionization_model_residuals} we show the residuals between the two, which span the range of -0.2 to 0.2 dex. The standard deviation of the residuals is $\sigma = 0.11$ dex, which we adopt as the uncertainty of the analytic expression. 

To test the robustness of our analytical expression, and in particular the dependence of line ratios on the gas density, we considered a more complicated model, consisting of four geometrically-thin, optically thick, gas clouds. All gas clouds have solar metallicity and ISM-type grains. The central source is characterized by a mean energy of an ionizing photon of 2.56 Ryd. We varied the distance of each gas cloud and its hydrogen density, such that all models have the same ionization parameter of $\log U = -2.75$. We set the hydrogen density in the clouds to be $\log n_{\mathrm{H}} (\mathrm{cm^{-3}}) = 5$, $\log n_{\mathrm{H}} (\mathrm{cm^{-3}}) = 4$, $\log n_{\mathrm{H}} (\mathrm{cm^{-3}}) = 3$, and $\log n_{\mathrm{H}} (\mathrm{cm^{-3}}) = 2$. 

As a first test, we examine the emission line ratios [OIII]/H$\beta$ and [NII]/H$\alpha$ of each individual cloud. Since the different clouds were defined to have the same ionization parameter, we expect their line ratios to be similar. The resulting line ratios differ from each other by less than 0.1 dex, supporting our suggestions that the hydrogen density does not affect the observed line ratios, for a constant ionization parameter. We used these line ratios and equation \ref{eq:5} to predict the ionization parameter, and found $\log U$ of -2.68, -2.78, -2.87, -2.96 respectively. These values differ from the input ionization parameter, $\log U = -2.75$, by at most $\sim $0.2 dex, consistent with the residuals shown in figure \ref{f:photoionization_model_residuals}. 

Next, we consider the case of a combination of clouds with different densities and distances. We normalize the four models to have a similar covering factor. This is similar to using the mean of the four previous ionization parameters. We extracted the predicted emissivities of [OIII], [NII], H$\alpha$, and H$\beta$ in the four different clouds, and summed them to obtain the total emissivity in each of the lines for the model. We then used the total [OIII]/H$\beta$ and [NII]/H$\alpha$ line ratios, and estimated the ionization parameter using the analytical expression from equation \ref{eq:5}. The predicted ionization parameter is $\log U = -2.83$. The difference between the input ionization parameter and the predicted is 0.08 dex, which is smaller than our adopted uncertainty. This suggests that the simple analytical form in equation \ref{eq:5} is expected to hold for more complex scenarios, where the outflow is stratified, with a large range of density and location.

\end{document}